\documentclass[letterpaper,twocolumn,english,preprintnumbers,amsmath,amssymb,superscriptaddress,footinbib,
prx
]{revtex4-2}
\usepackage{graphicx} 
\usepackage{silence}
\usepackage{amsthm}
\usepackage{amssymb}
\usepackage{xcolor}
\usepackage[colorlinks = true, citecolor=blue, linkcolor=black,urlcolor=purple]{hyperref}
\usepackage{enumitem}
    \setlist[itemize]{noitemsep, topsep=0pt}
    \setlist[enumerate]{noitemsep, topsep=0pt}
\usepackage{verbatim}
\usepackage{mathtools,amssymb,amsmath}
\usepackage{bm}
\usepackage{yhmath}
\usepackage[inline]{asymptote}
\usepackage{cancel}
\usepackage{relsize}
\usepackage{array}
\usepackage{bm}
\usepackage{float}

\newcommand{\ket}[1]{|#1\rangle}

\newcommand{\be}{\begin{equation}}
\newcommand{\ee}{\end{equation}}

\newcommand{\Tr}{{\rm Tr\,}}

\usepackage[normalem]{ulem}
\usepackage{color}

\usepackage{afterpage}

\newcommand{\ibmyorktown}{IBM Quantum, IBM T.J. Watson Research Center, Yorktown Heights, 10598, USA}
\newcommand{\ibmcambridge}{MIT-IBM Watson AI Lab,  Cambridge MA, 02142, USA}
\newcommand{\Duke}{Duke Quantum Center, Duke University, Durham, NC 27701, USA}
\newcommand{\Dukeelec}{Department of Electrical and Computer Engineering, Duke University, Durham, NC 27708, USA}
\newcommand{\HarvardPhysics}{Department of Physics, Harvard University, Cambridge, MA 02138, USA}
\newcommand{\HarvardHQI}{Harvard Quantum Initiative, Harvard University, Cambridge, MA 02138, USA}
\newcommand{\Stanford}{Department of Physics, Stanford University, Stanford, CA 94305, USA}
\newcommand{\UFFI}{ Instituto de F\'isica, Universidade Federal Fluminense,  Niter\'oi, RJ, 24210-346, Brazil}

\usepackage{algorithm}
\usepackage{algpseudocode}

\makeatletter
\newsavebox{\@brx}
\newcommand{\llangle}[1][]{\savebox{\@brx}{\(\m@th{#1\langle}\)}
  \mathopen{\copy\@brx\kern-0.5\wd\@brx\usebox{\@brx}}}
\newcommand{\rrangle}[1][]{\savebox{\@brx}{\(\m@th{#1\rangle}\)}
  \mathclose{\copy\@brx\kern-0.5\wd\@brx\usebox{\@brx}}}
\makeatother

\usepackage{xcolor}

\usepackage{tikz} 
\usetikzlibrary{quantikz} 
\usepackage{adjustbox} 
\usepackage{bibunits} 
\newcommand{\ourtitle}{Measuring central charge on a universal quantum processor}
\makeatletter 
\renewcommand{\fnum@figure}{\textbf{Fig.~\thefigure}}
\def\@caption@fignum@sep{\textbf{.} }
\makeatother 

\begin{document}
\makeatother

\title{\ourtitle}
\author{Nazl\i \ U\u{g}ur K\"oyl\"uo\u{g}lu$^{*\dag}$}
\affiliation{\ibmyorktown}
\affiliation{\HarvardPhysics}
\affiliation{\HarvardHQI}
\affiliation{\Stanford}
\author{Swarndeep Majumder$^{*\ddagger}$}
\affiliation{\ibmyorktown}
\affiliation{\Duke}
\affiliation{\Dukeelec}
\altaffiliation{These authors contributed equally to this work.\\
$^\dag$nazliugurkoyluoglu@g.harvard.edu\\
$^\ddagger$swarnadeep.majumder@ibm.com\\
$^\&$knajafi@ibm.com}
\author{Mirko Amico}
\affiliation{\ibmyorktown}
\author{Sarah Mostame}
\affiliation{\ibmyorktown}
\author{Ewout van den Berg}
\affiliation{\ibmyorktown}
\author{M. A. Rajabpour}
\affiliation{\UFFI}
\author{Zlatko Minev}
\affiliation{\ibmyorktown}
\author{Khadijeh Najafi$^\&$}
\affiliation{\ibmyorktown}
\affiliation{\ibmcambridge}

\maketitle

\textbf{Central charge is a fundamental quantity in conformal field theories (CFT), and plays a crucial role in determining universality classes of critical points in two-dimensional systems. Despite its significance, the measurement of central charge has remained elusive thus far. In this work, we present the first experimental determination of the central charge using a universal quantum processor. Using a classically optimized variational quantum circuit and employing advanced error mitigation techniques, we successfully prepare ground states of various $1+1D$ quantum spin chain models at their critical point. Leveraging the heavy-hex structure of IBM quantum processors, we are able to implement periodic boundary conditions and mitigate boundary effects. We then extract the central charge from the scaling behavior of the sub-leading term of R{\'{e}}nyi generalizations of classical Shannon entropy, computed for local Pauli measurements in the conformal bases ($\sigma^{z}$ and $\sigma^x$). The experimental results are consistent with the known central charge values for the transverse field Ising (TFI) chain ($c=0.5$) and the XXZ chain ($c=1$), achieving relative errors as low as 5 percent.

}
\vspace{0.15cm}

Conformal symmetry, i.e. invariance under angle-preserving transformations of spacetime, is essential to the analysis of critical phenomena and high-energy physics. For instance, short-range interacting systems at the phase transition point are hypothesized to exhibit conformal symmetry~\cite{Polyakov}, which offers insights into their scale invariant properties. In high-energy physics, particularly in string theory, conformal symmetry is an important property of the worldsheet~\cite{Polchinski}, the 
surface traced out by a string as it moves through spacetime. 
In two dimensions, the conformal group has infinite generators, which allows it to characterize a system's universal physical properties based solely on this symmetry~\cite{Belavin1984}.
The associated Virasoro algebra~\cite{Virasoro} has a key parameter, the central charge. 
In fact, the value of the central charge, combined with the coefficients from the primary fields' operator product expansion (OPE), identifies the system's universality class, see~\cite{DiFrancesco_cft} for a review.

The central charge manifests itself in numerous physical quantities, including correlation functions of the energy-momentum tensor and scaling of the system's free energy density~\cite{BCN1986,Affleck}. It also appears in the scaling behavior of information theoretic measures like entanglement (von Neumann-R{\'{e}}nyi) entropies~\cite{Holzhey_1994} and classical (Shannon-R{\'{e}}nyi) entropies~\cite{Alcaraz_2013,Stephan_2014B}. 
Despite being a fundamental parameter in various physical quantities associated with two dimensional conformal field theories (CFTs),
the experimental determination of central charge has remained elusive over the past 40 years.
A significant challenge in using pre-quantum information metrics 
was the concurrent need to determine the sound velocity in the system~\cite{BCN1986,Affleck}. On the other hand, information theoretic metrics like entanglement or classical entropies do not face this issue. At the critical point, these quantities exhibit 
either primary or secondary logarithmic behavior as a function of subsystem size~\cite{Holzhey_1994,Vidal_2003,Alcaraz_2013,Stephan_2014B}, with a scaling coefficient that has direct dependence on the central charge;
for more information see Supplementary material~\ref{SI:CFT}.

\begin{figure*}[t!]
\setlength{\tabcolsep}{10pt}
\centering
\includegraphics[width=1.0\textwidth]{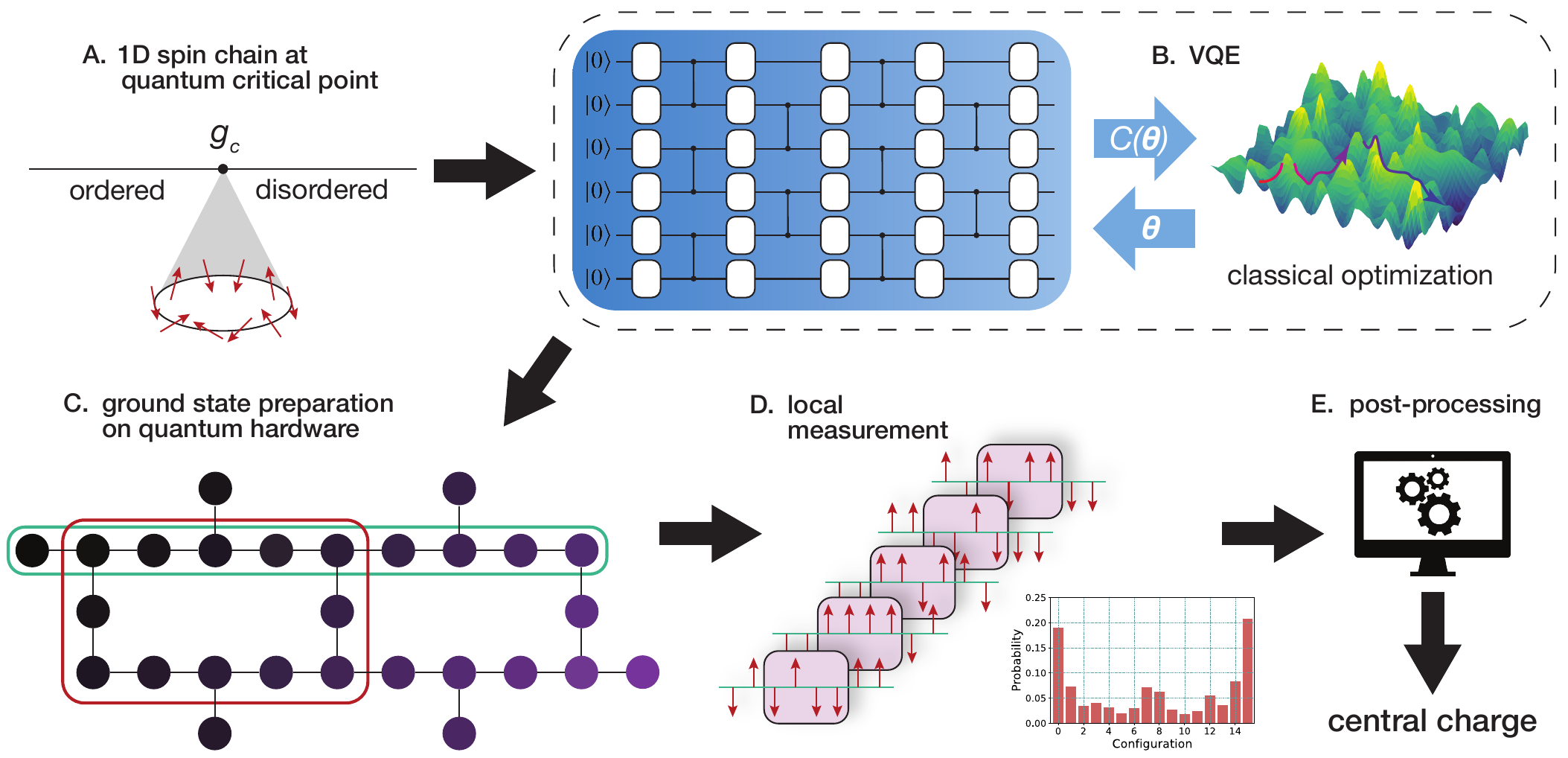}
\caption{\textbf{Measuring central charge from local observables.} (\textbf{A}) Illustration of quantum phase transition for generic quantum spin chain models $H(g)$ as a function of tuning parameter $g$; $g_c$ indicates the quantum critical point. (\textbf{B}) Simulated VQE algorithm where the parameters $\mathbf{\theta}$ of a checkerboard ansatz are classically optimized to minimize energy expectation $C(\mathbf{\theta})$ with respect to a quantum spin chain Hamiltonian; the resulting quantum circuit is then used to experimentally prepare the ground state. (\textbf{C}) Selection of qubits on the heavy-hex layout of IBM quantum processors, to construct spin chains of size $L=10$ with open boundary conditions (green box) and size $L=12$ with periodic boundary conditions (red box). (\textbf{D}) Spin configuration probabilities obtained from single-shot measurements in a local basis. Inset: $\sigma^{z}$-basis bitstring probabilities measured from the experimentally prepared ground state of the critical TFI Hamiltonian, presented for a $L=4$ chain for clarity. (\textbf{E}) Classical post processing for extracting the central charge, by fitting the Shannon-R\'enyi entropies of the measured configuration probabilities.
}
\label{fig1:Experimental set-up}
\end{figure*}

Thus, to experimentally measure the central charge,
two key challenges must be addressed: successful preparation of the critical ground state on hardware and efficient measurement of the aforementioned quantities. 
Significant progress has been made in  
utilizing 
quantum processors with hundreds of qubits 
to explore a wide class of dynamical quantum many-body phenomena, such as many body localization~\cite{shtanko2023uncovering}, thermalization and critical dynamics~\cite{keesling2019,KZ2024_Zurich,andersen2024thermalization,manovitz2024quantumcoarseningcollectivedynamics,Chen:2023tfg}, time crystals~\cite{Mi_2021}, and gauge theories~\cite{Gague2023}, as well as to study exotic phases of matter and topological order~\cite{Semeghini2021,Minev2024,xu2024nonabelian,Satzinger_2021,Harle_2023}. However, realizing and probing high-fidelity ground states at quantum phase transition points with digital circuits remain more challenging~\cite{Dupont_2022,Dborin_2022}.

In this work, we experimentally prepare $1+1D$ quantum critical ground states (see Fig.~\ref{fig1:Experimental set-up}(A)) and extract the central charge for two distinct universality classes: the transverse-field Ising (TFI) model with $\mathbb{Z}_2$ symmetry, and the XXZ model with $U(1)$ symmetry. Additionally, we explore the extension of our approach to the newly introduced tricritical spin chain with supersymmetry~\cite{O'Brian2018} (see Supplementary material~\ref{SI:Tricritical}). Our experiments are performed on the 27-qubit IBM Falcon processor \textit{ibm\_peekskill} and 65-qubit Hummingbird processor \textit{ibm\_ithaca}; detailed information about the hardware and the selection of qubits can be found in Supplementary material~\ref{SI:qubit_choice}.

For ground state preparation, we employ classically optimized variational quantum circuits. In particular, we numerically simulate the variational quantum eigensolver (VQE) algorithm to optimize a checkerboard ansatz $|\psi(\boldsymbol{\theta})\rangle$, which consists of single-qubit rotations parameterized by angles  $\boldsymbol{\theta}$, interleaved with alternating layers of two-qubit gates implementing nearest-neighbor interactions (see Fig.~\ref{fig1:Experimental set-up}(B)). The objective is to minimize a cost function defined as energy expectation with respect to the target Hamiltonian $H$:       $C(\boldsymbol{\theta})=\langle\psi(\boldsymbol{\theta})|H|\psi(\boldsymbol{\theta})\rangle$~\cite{Peruzzo_2014,Cerezo_2021}. 
We execute the resulting quantum circuit $|\psi(\boldsymbol{\theta}_\mathrm{opt})\rangle$ to prepare the target ground state of $H$ on a chain of qubits isolated on the hardware, as shown in Fig.~\ref{fig1:Experimental set-up}(C). We note that the checkerboard ansatz was shown to accurately encode ground states of quantum critical chains as the number of layers scales linearly with system size~\cite{Bravo_Prieto_2020}. 
Supplementary material~\ref{SI:VQE} details the number of layers we implement, our choice of classical optimization algorithm, and the number of iterations needed for convergence. 

\begin{figure*}[t]
\setlength{\tabcolsep}{10pt}
\centering
\includegraphics[width=1.0\textwidth]{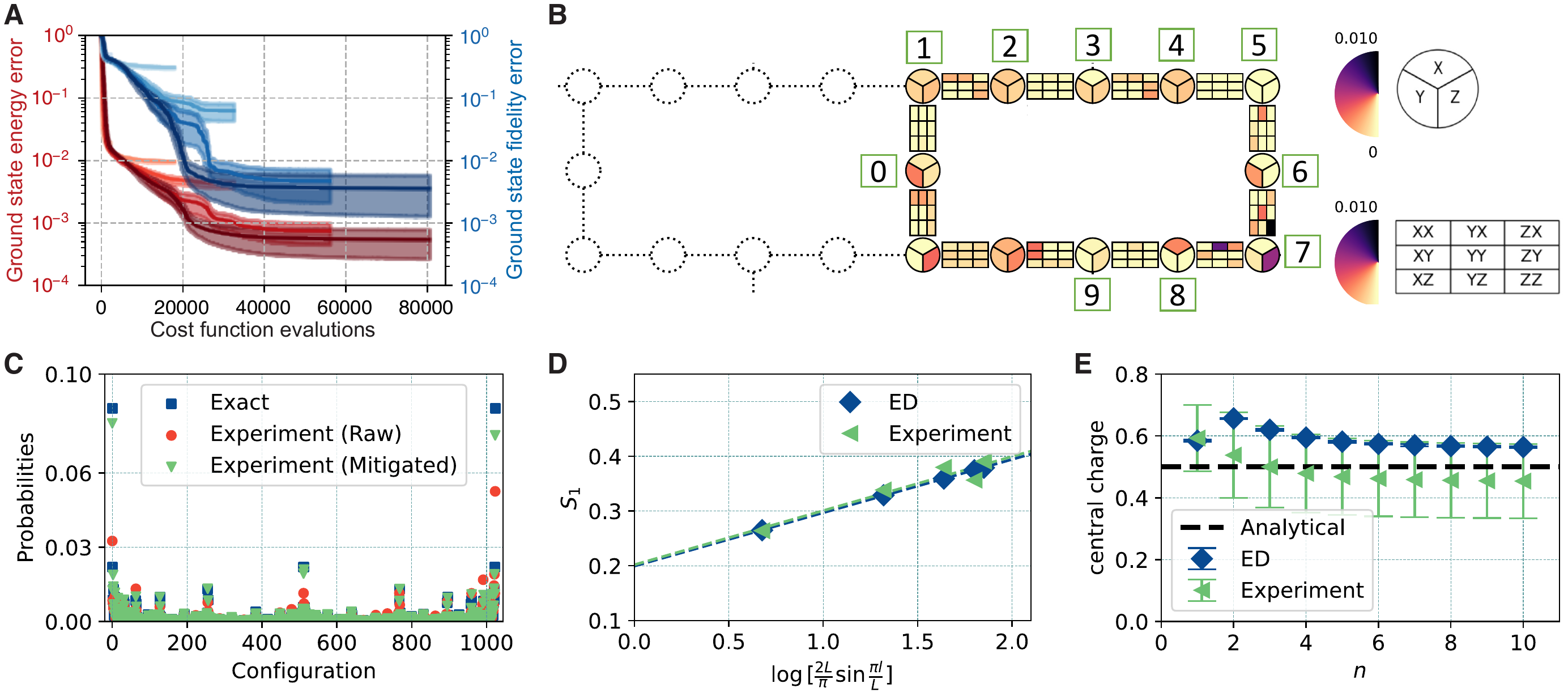}
\caption{\textbf{Measuring entanglement entropies on the critical TFI chain using probabilistic error cancellation.}
(\textbf{A}) 
Numerically simulated VQE algorithm 
for preparing the ground state of the TFI model at the critical point ($h/J=1$), on an open spin chain of size $L=10$, 
achieving lowest (theoretical) relative energy error of $10^{-3}$ (red) and state infidelity of $10^{-2}$ (blue) at convergence. (\textbf{B}) Visualization of the sparse Pauli noise model generator coefficients used for PEC on a $27$-qubit IBM Falcon processor. Circles represent qubits, with three internal wedges corresponding to single $X$, $Y$, and $Z$ Lindbladian coefficients. The $3\times 3$ matrices represent two-body coefficients 
corresponding to different possible pairs. (\textbf{C}) $\sigma^{z}$-basis bitstring probabilities computed from raw experimental data (red) as well as error-mitigated data via PEC (green), compared with exact diagonalization (blue). The wavefunction reconstructed from error-mitigated probabilities achieves a state preparation fidelity of $F=|\langle \psi_0|\psi_{\mathrm{exact}}\rangle|^2=0.96$ on the hardware. (\textbf{D}) von Neumann entanglement entropy $S_1$ as a function of subsystem size $l$, computed for wavefunctions recontstructed from $\sigma^z$-basis bitstring probabilities, showing good agreement between error-mitigated experimental data (green) and exact diagonalization (blue). 
The value of central charge is obtained from the slope of the fitted line. (\textbf{E}) Central charge values obtained from fits to R\'enyi entanglement entropies $S_n$ of different moments $n=1,...,10$, computed from mitigated experimental data as described in (D), compared against fits to exact diagonalization results and the analytical value of $c=0.5$.
The most accurate value $c=0.49\pm0.13$ is obtained for $n=3$. 
}
\label{fig2:TFIM-OP}
\end{figure*}

\begin{figure*}[t]
\setlength{\tabcolsep}{10pt}
\centering
\includegraphics[width=0.98\textwidth]{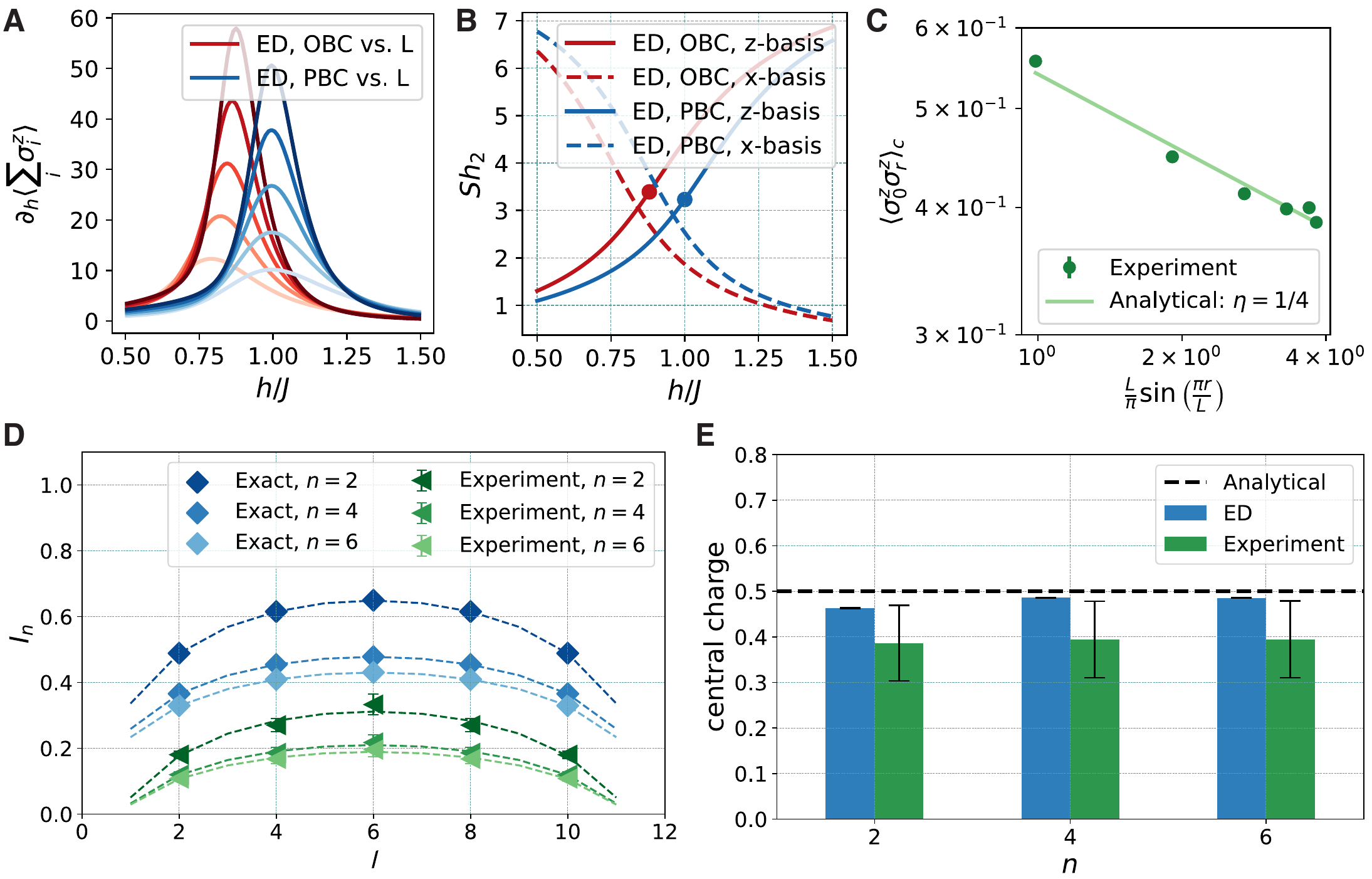}
\caption{
\textbf{Extracting central charge from Shannon entropies of the critical TFI chain with periodic boundary conditions.} (\textbf{A}, \textbf{B}) Finite size effects on the critical point analyzed using exact diagonalization. The derivative of the order parameter peaks at a value of $h/J$ closer to the thermodynamic critical point $h_c/J=1$ and less sensitive to system size ($L=4,6,8,10,12$ shown with increasing shade) in the presence of periodic boundary conditions (blue), compared to open boundary conditions (red). Similarly, the second-moment Shannon-R\'enyi entropies of $\sigma^z$ and $\sigma^x$-basis bitstring probabilities, computed for a $L=12$ chain, satisfy the expected relation $Sh^z_2-Sh^x_2=\ln{2}$ (indicated by marker) closer to $h_c/J=1$ in the presence of PBC (blue), compared to OBC (red). (\textbf{C}) Two-point $\sigma^z$-connected correlators obtained from the experimentally prepared ground state of the critical TFI chain ($h/J=1$) on a $L=12$ periodic chain, consistent with the predicted power-law decay with analytical exponent $\eta=1/4$ (green line). (\textbf{D}) R\'enyi difference measure $I_n$ calculated for moments $n=2,4,6$ via Eq.~\eqref{eqn:mutual_info}, using $\sigma^{x}$-basis configuration probabilities as a function of subsytem size $l$. Experimental results (green) show similar dependence on subsystem size as exact diagonalization (blue), with a relative offset. (\textbf{E}) Central charge extracted from fitting $I_n$ to Eq.~\eqref{eqn:mutual_l}, resulting in estimates of $c=0.39\pm0.08$ for $n=2,4,6$, compared to the analytical value of $c=0.5$. 
}
\label{fig:TFIM_Periodic}
\end{figure*}

Next, we address the measurement of information theoretic quantities, which are needed to extract the central charge. Consider the ground state $|\psi_0\rangle$ of a quantum spin chain of size $L$ (see Fig.~\ref{fig1:Experimental set-up}(A)). 
The ubiquitous von Neumann-R\'enyi entanglement entropies~\cite{Amico:2007ag,Eisert2010,Laflorencie:2016,Abanin_2019,vermersch2023manybody} of moment $n>0$, for a subsystem of $l$ sequential spins, reads:
\begin{equation}
S_n(l) = \frac{1}{1-n} \ln \operatorname{Tr} \rho_l^n,
\label{eqn:ent_ent}
\end{equation}
where $\rho_l={\rm Tr}_{L-l}|\psi_0\rangle\langle \psi_0|$ is the reduced density matrix associated with the subsystem.
At the critical point, the ground state is described by a CFT, 
and we have the following relation for a subsystem of size $l$ selected from one end of an open chain~\cite{Calabrese_2004}:
\begin{equation}
S_n(l) = \frac{c}{12}\frac{n+1}{n}\ln \left(\frac{2L}{\pi a}\sin{\frac{\pi l}{L}}\right)+\mathcal{O}(1), 
\label{eqn:CFT-EE-PBC-main}
\end{equation}
where $c$ is the central charge and $a$ is the lattice constant. 
Entanglement entropy measurements have been demonstrated experimentally, either through swap operators~\cite{Islam:2015ecs,bluvstein2022} or randomized measurements~\cite{BW_Zoller2023,Roos:2019}. 

On the other hand, one can sidestep entanglement entropies and instead compute classical Shannon-R\'enyi entropies, relying solely on bitstring probabilities that are generically accessible from local projective measurements, see Fig.~\ref{fig1:Experimental set-up}(D). 
More precisely, the ground state can be expressed in some local basis as $|\psi_0\rangle=\sum_{\nu}a_{\nu}|\nu\rangle$, where $p_{\nu}=|a_{\nu}|^2$ corresponds to the probability of measuring bitstring $\nu$ in this basis. These $p_{\nu}$ can be used to compute marginal probabilities $p_{\bar{\nu}}$ associated with configurations $\bar{\nu}$ of a subsystem of size $l$, and obtain $n^\mathrm{th}$-moment Shannon-R\'enyi entropies of this subsystem:
\begin{equation} 
Sh_n(l)=\frac{1}{1-n}\ln \sum_{\bar{\nu}} p_{\bar{\nu}}^n\;.
\label{eqn:Shannon_L}
\end{equation}
Typically, these quantities scale proportionally with subsystem size. However, a logarithmic sub-leading term emerges for critical systems (see Supplementary material~\ref{SI:CFT}). Thus, it is often advantageous to eliminate the linear term and work directly with the following quantity:
\begin{equation} 
I_n(l,L-l)=Sh_n(l)+Sh_n(L-1)-Sh_n(L),
\label{eqn:mutual_info}
\end{equation}
which we refer to as the R\'enyi difference measure (RDM).
For $n=1$, the RDM is called mutual Shannon entropy, which has a physical interpretation as the amount of information gained about subsystem $L-l$ by measuring subsystem $l$. $I_n(l,L-l)$ for other moments $n$ lack such interpretation, as the quantity can be negative in some cases~\cite{Teixeira2012,verdu2015}. 
We emphasize that while there are an exponential number of configurations, the scaling of the quantity $I_n(l,L-l)$ is typically dominated by the largest probabilities, which can be accurately estimated using a moderate number of projective measurements.

Although Shannon-R\'enyi entropies are basis-dependent, it has been shown that there are particular bases whose finite-size scaling of $I_n(l,L-l)$ can be related to the central charge of the underlying CFT.
In such bases, referred to as conformal bases in~\cite{MAR2014}, the RDMs computed for periodic chains satisfy:~\cite{Alcaraz_2013,Stephan_2014B,MAR2014}
\begin{equation} 
I_n(l,L-l)=\frac{c_n}{4}\ln\left[\frac{L}{\pi}\sin\left(\frac{\pi l}{L}\right)\right]+\gamma_n.
\label{eqn:mutual_l}
\end{equation}
Contrary to entanglement entropies, the coefficient $c_n$ in Eq.~\eqref{eqn:mutual_l} has a discontinuous behavior.  For systems with $U(1)$ symmetry, such as the XXZ chain, $c_n=c$ for $n<n_t$, and $c_n=c\frac{n}{n-1}$ for $n>n_t$, where $n_t$ is dependent on the model and the basis. For systems with $\mathbb{Z}_n$ symmetry, such as the TFI and Potts models, $c_n=c\frac{n}{n-1}$ for $n>n_t$, where $n_t$ is typically around $1$~\cite{MAR2014}. 

Given the outlined protocols for critical ground state preparation and central charge measurement, we now consider the transverse-field Ising Hamiltonian, defined as 
\begin{eqnarray}\label{eq:XY hamiltonian}\
H_{\mathrm{TFI}}=-J\sum_{l=1}^{L}\sigma_l^z\sigma_{l+1}^z-h\sum_{l=1}^L\sigma_l^x\;,
\end{eqnarray}
where $\sigma^{\alpha}$ for $\alpha\in\{x,y,z\}$ denote the Pauli operators, $J$ is the coupling strength between neighboring spins, and $h$ is the strength of the external transverse magnetic field. At the critical point $h_c/J=1$, the ground state is described by a CFT with a central charge of $c=1/2$. As shown in Fig.~\ref{fig2:TFIM-OP}(A), the simulated VQE algorithm converges to a quantum circuit that can theoretically prepare this ground state with a fidelity of 
$\mathcal{F}_{\mathrm{ideal}}=|\langle \psi(\theta)|\psi_{\mathrm{exact}}\rangle|^2=0.99$. However, experimental noise poses a significant challenge to ground state preparation,
necessitating the use of advanced error mitigation schemes.

Our error mitigation strategy consists of probabilistic error cancellation (PEC) based on a sparse Pauli-Lindblad noise model~\cite{berg2022probabilistic}, combined with model-free readout error mitigation~\cite{PhysRevA.105.032620}. PEC involves learning the coefficients of single and two-qubit Pauli errors for various  layers of  parallel two-qubit gates within the circuit, as depicted in Fig.~\ref{fig2:TFIM-OP}(B).
The noise cancellation is performed by quasi-probabilistic sampling of circuit instances,
and the error-mitigated observable expectation value is obtained as a weighted average of the estimates from the various sampled circuits; for more details see Supplementary material~\ref{SI:error mitigation}. In order to obtain the error-mitigated bitstring probabilities, 
we perform error mitigation on all Pauli expectation values in the chosen basis, and subsequently apply the Walsh-Hadamard (WH) transform. 
We note that PEC applies an inverse noise map through quasi-probabilistic sampling, which can sometimes result in unphysical expectation values 
and lead to negative probabilities after the WH transform, particularly for configurations with extremely low probabilities. To address this issue, we discard probabilities below a certain small threshold and subsequently renormalize the overall probability distribution, see Supplementary material~\ref{SI:Boostrapping}. In our experiments, we execute $10,000$ different error-mitigated circuits, each consisting of 200 shots. In Fig.\ref{fig2:TFIM-OP}(C), we compare $\sigma^z$-basis bitstring probabilities $p_z$ obtained from hardware against exact diagonalization, and highlight the improvement achieved through error mitigation. 

Since the TFI chain Hamiltonian is stoquastic~\cite{bravyi2007}, i.e. has real and non-positive off-diagonal elements in the computational basis, we are able to successfully reconstruct the ground state from the mitigated $\sigma^z$-basis bitstring probabilities in post-processing, assuming a coherent pure state: $|\psi_0\rangle=\sum_{z}\sqrt{p_{z}}|z\rangle$, which is consistent with a fidelity of $\mathcal{F}_\mathrm{exp}=|\langle \psi_0|\psi_{\mathrm{exact}}\rangle|^2=0.96$ with respect to the exact critical ground state (note that we have not verified the coherences of the state). 
We thus demonstrate the first experimental determination of the probability amplitudes of a high fidelity quantum critical ground state on a digital quantum processor. 

With access to the quantum state, we directly compute the von Neumann-R\'enyi entanglement entropies and determine the central charge by analyzing their scaling with subsystem size, as provided in Eq.~\ref{eqn:CFT-EE-PBC-main}. In Fig.~\ref{fig2:TFIM-OP}D, we demonstrate that the von-Neumann entanglement entropies calculated from experimental data closely match exact diagonalizaiton, and yield an estimate of $c=0.59\pm0.11$ for the central charge. 
In  Fig.~\ref{fig2:TFIM-OP}(E), we apply this method to R\'enyi entanglement entropies of higher moments $n$, and obtain the most accurate value $0.50\pm0.13$ for $n=3$. We report central charge estimates for various $n$ in Supplementary material~\ref{SI:Boostrapping}. 

\begin{figure*}[t]
\setlength{\tabcolsep}{10pt}
\includegraphics[width=0.98\textwidth]{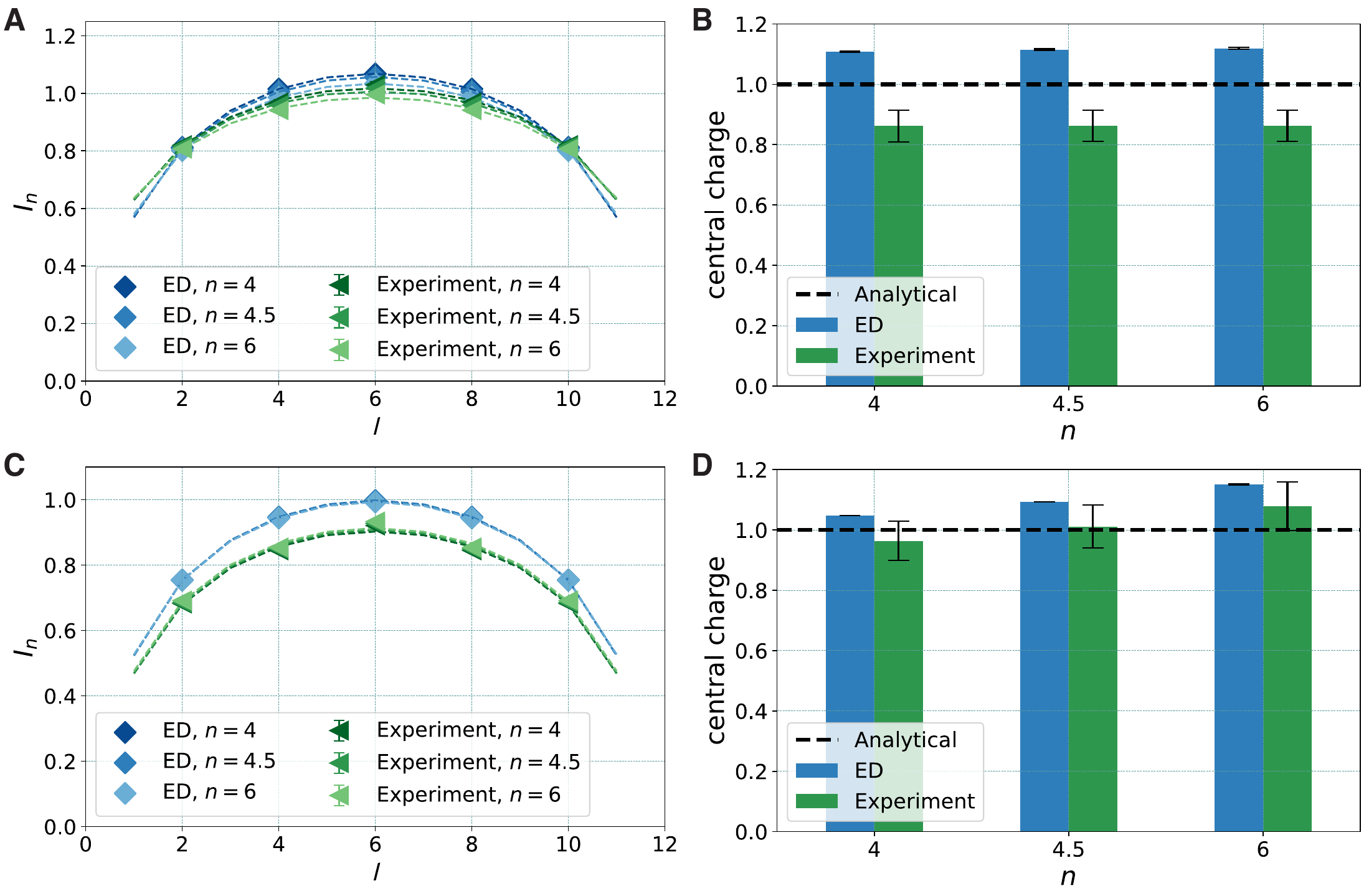}
\caption{\textbf{Generalizaiton of protocol to XXZ chain with periodic boundary conditions.} (\textbf{A}) R\'enyi difference measure $I_n$ calculated for moments $n=4,4.5,6$ via Eq.~\eqref{eqn:mutual_info}, using $\sigma^{x}$-basis configuration probabilities as a function of subsytem size $l$, obtained from the experimentally prepared ground state of the critical XXZ chain ($\Delta=-0.5$) on a $L=12$ periodic chain. Experimental results (green) show similar dependence on subsystem size as exact diagonalization (blue), with a relative offset. (\textbf{B}) Central charge extracted from fitting $I_n$ to Eq.~\eqref{eqn:mutual_l}, resulting in estimates of $c=0.86\pm0.05$ for $n=4,4.5,6$, compared to the analytical value of $c=1$. 
(\textbf{C}) $I_n$ calculated for moments $n=4,4.5,6$ in the $\sigma^{z}$-basis. (\textbf{D}) This basis yields more accurate estimats of the central charge, $c=0.96\pm0.07$,$1.01\pm0.07,1.08\pm0.08$ for $n=4,4.5,6$, respectively. 
}
\label{fig:XXZ-Periodic}
\end{figure*}

In order to mitigate boundary effects on the central charge measurement, we then investigate the critical TFI chain with periodic boundary conditions (PBC). As shown in Fig.~\ref{fig:TFIM_Periodic}(A-B), our numerical analyses of the derivative of the order parameter, as well as the $n=2$ Shannon-R\'enyi entropy in local bases, as a function of $h/J$ across different system sizes confirm that PBC is more robust against system size effects than open boundary conditions (OBC). 
Based on these findings, we experimentally prepare the ground state of a periodic spin chain of length $L=12$, following the procedure outlined for the open chain. Then, we perform projective measurements in the $\sigma^z$-basis and extract error-mitigated bitstring probabilities. In Fig.~\ref{fig:TFIM_Periodic}(C), we show that the $z$-basis connected correlators $\langle \sigma^z_0 \sigma^z_r\rangle_c=\langle \sigma^z_0 \sigma^z_r\rangle-\langle \sigma^z_0 \rangle\langle\sigma^z_r\rangle$ obtained from experiment are consistent with the expected power-law decay with analytical critical exponent $\eta=1/4$. To determine the central charge, we calculate the classical Shannon-R\'enyi entropies $Sh_n$ of subsystem configuration probabilities (instead of using entanglement entropies). The $\mathbb{Z}_2$ symmetry of the TFI model imposes constraints on $\sigma^x$-basis measurements, namely the elimination of configurations with odd parity in the ground state, which we exploit in post-processing of the error-mitigated bitstring probabilities (see Supplementary material~\ref{SI:basis choice},~\ref{SI:Boostrapping}). 
In Fig.~\ref{fig:TFIM_Periodic}(D), we use these probabilities to obtain the RDMs $I_n(l,L-l)$ introduced in Eq.~\eqref{eqn:mutual_info} for various moments $n$, and extract the central charge from their scaling following Eq.~\eqref{eqn:mutual_l}. 
As a result, we estimate $c=0.39\pm0.08$ for $n=2,4,6$, as shown in Fig.~\ref{fig:TFIM_Periodic}(E). We report error bars by propagating random errors arising from projection noise and error mitigation (evaluated via bootstrapping), as well as systematic errors from the fitting procedure, for more details see Supplementary material~\ref{SI:Boostrapping}. In comparison, a similar analysis for exact diagonalized configuration probabilities yield $c=0.463\pm0.002,0.486\pm0.001,0.485\pm0.001$ for $n=2,4,6$, where error bars account for systematic errors from the fitting procedure. 

Extending beyond the TFI chain, we successfully apply our methodology to other models, namely the XXZ Hamiltonian on a $L=12$ chain with periodic boundary conditions:
\begin{eqnarray}
H_{\mathrm{XXZ}}=-\sum_{l=1}^{L}\Big{(}\sigma_l^x\sigma_{l+1}^x+\sigma_l^y\sigma_{l+1}^y+\Delta\sigma_l^z\sigma_{l+1}^z\Big{)}\;, 
\label{eqn:xxz}
\end{eqnarray}
parameterized by anisotropy $\Delta$. Within the critical regime $-1\leq \Delta <1$, the ground state is gapless and described by the  compactified boson CFT, with a central charge of $c=1$~\cite{Giamarchi}. Specifically, we focus on antiferromagnetic Ising interactions with $\Delta = -0.5$, for which the simulated VQE algorithm converges to higher fidelities.
Using the optimized quantum circuit, we follow a similar protocol for ground state preparation, error mitigation, and post-processing as for the periodic TFI chain. In this case, we note that the RDMs $I_n(l,L-l)$ exhibit discontinuities at $n_t^z = 3$ and $n_t^x = \frac{4}{3}$, and are particularly robust to finite size effects at higher moments $n$~\cite{MAR2014}.
Therefore, we utilize 
$n = 4$, $4.5$ and $6$. From analysis of configuration probabilities in the $\sigma^x$ basis, we obtain central charge estimates of $c=0.86 \pm 0.05$, as shown in Fig.~\ref{fig:XXZ-Periodic}(A,B) (compared to best estimate $c=1.108 \pm 0.002$ obtained from exact diagonalization). On the other hand, in the $\sigma^z$-basis, the $U(1)$ symmetry of the model restricts the occurrence of certain bitstrings, as the ground state is an eigenstate of total $\sigma^z$-magnetization, specifically a half-filled state. By leveraging this property in post-processing (see Supplementary material~\ref{SI:basis choice},~\ref{SI:Boostrapping}), we are able to obtain more accurate central charge estimates in this basis: $c=0.96\pm 0.07 , 1.01\pm 0.07,1.08\pm 0.08 $ for $n=4,4.5,6$ respectively, as shown in Fig.~\ref{fig:XXZ-Periodic}(C-D) (compared to $c=1.0477 \pm 0.0003,1.0932 \pm 0.0004,1.151\pm 0.001$ obtained from exact diagonalization). 
These values of the central charge are remarkably accurate, with relative error as low as 5$\%$, especially considering the small system size, limited number of projective measurements, and various uncertainties associated with the probability extraction and error mitigation processes.

In conclusion, our work presents the first experimental measurements of the central charge associated with various conformal field theories,
through a novel, generic scheme that requires only local measurements and is independent of the Hamiltonian. Moreover, by leveraging the heavy-hex topology of IBM quantum processors, we study spin chains with both open and periodic boundary conditions, effectively minimizing boundary effects in the latter. With the implementation of advanced error mitigation techniques such as PEC, we successfully extract central charge values with relative errors as low as 5 percent for the XXZ chain using $\sigma^z$-basis measurements. 

The theoretical analysis of central charge is inherently challenging. For instance, while certain scaling relations for entanglement entropy at critical points have been established (see Eq.\ref{eqn:CFT-EE-PBC-main}), general relations for different boundary conditions remain largely unknown. On the other hand, these results highlight the potential of universal quantum processors as promising tools for determining the central charge experimentally, which could significantly enhance the study of CFTs. To this end, we also explore the application of our protocol to the tricritical Ising model, recently introduced in~\cite{O'Brian2018}, which manifests supersymmetry (see Supplementary material~\ref{SI:Tricritical}); however, our results are limited by small system size.
In particular, although we succeed in experimentally preparing a high-fidelity ground state on an $L=10$ open chain, the simulated VQE algorithm fails to converge on longer periodic chains due to long-range interactions, underscoring the need for scalable ground state preparation methods.
Scaling variational methods to larger system sizes would require evaluating the cost function at each iteration directly on the hardware instead. Additionally, the achievable performance remains unclear due to the existence of barren plateaus~\cite{larocca2024review}. Therefore, in Supplementary material~\ref{SI:direct_state_prep}, we discuss alternative methods that could be employed for ground state preparation in the future~\cite{verstraete2009quantum}. Moreover, we highlight the tremendous developments in mapping tensor network states onto quantum circuits~\cite{haghshenas2023probingcriticalstatesmatter,smith2024constantdepth,Igancio2024}. These developments could pave the way for studying critical states on larger systems and extracting the central charge across a wide range of critical quantum spin chains in the near future.

\textbf{Acknowledgements.} We acknowledge stimulating discussions with Francesco Alcaraz, Sergey Bravy, Andi Gu, Jad Halimeh, Ivano Tavernelli, and Peter Zoller. We thank Dolev Bluvstein, Edward Chen, Andrew Eddins, Youngseok Kim, Brad Mitchell, and Oles Shtanko for their valuable comments on our manuscript. 
MAR thanks CNPq and FAPERJ (grant number E-26/210.062/2023) for partial support. 

\let\oldaddcontentsline\addcontentsline
\renewcommand{\addcontentsline}[3]{}
\bibliography{references}
\let\addcontentsline\oldaddcontentsline

\clearpage
\pagebreak

\setcounter{page}{1}
\setcounter{equation}{0}
\setcounter{figure}{0}
\renewcommand{\theequation}{S.\arabic{equation}}
\renewcommand{\thefigure}{S\arabic{figure}}
\renewcommand*{\thepage}{S\arabic{page}}

\onecolumngrid

\begin{center}
{\large \textbf{Supplementary Materials for \\ ``\ourtitle"}}\\
\vspace{0.5cm}
Nazl\i \ U\u{g}ur K\"oyl\"uo\u{g}lu$^{1,2,3,4}, $Swarndeep Majumder$^{1,5,6}$, Mirko Amico$^{1}$, Sarah Mostameh$^{1}$\\
Ewout van den Berg$^{1}$, M. A. Rajabpour$^{7}$, Zlatko Minev$^{1}$,
Khadijeh Najafi$^{1,8}$\\ 
\vspace{0.25cm}
\textit{$^1$\ibmyorktown} \\
\textit{$^2$\HarvardPhysics}\\
\textit{$^3$\HarvardHQI}\\
\textit{$^4$\Stanford}\\
\textit{$^5$\Duke}\\
\textit{$^6$\Dukeelec}\\
\textit{$^7$\UFFI}\\
\textit{$^8$\ibmcambridge}\\
\end{center}

\tableofcontents

\vspace{1cm}

\twocolumngrid

\section{Theoretical analysis}
\label{SI:theoritical}

In this work, we lay the theoretical foundation for extracting the central charge from various physical quantities. We begin by introducing the fundamental concepts of conformal field theory, followed by a detailed discussion of the theoretical insights into emptiness formation probabilities, with particular emphasis on the emergence of central charge in the scaling behavior of this quantity. Subsequently, we explore the numerical extensions of the Shannon-Rényi entropy across different bases. Our review then extends to key studies on the transverse Ising model and the XXZ chain, where we also examine the impact of symmetries on measurement outcomes. We conclude this theoretical section with a discussion of the exact preparation of ground states and the application of the quantum Krylov method.
\subsection{Trace of central charge in quantum many body systems}
\label{SI:CFT}

Classical information theory relies on Shannon entropy and its extensions, known as the R\'enyi entropies. These entropies can be used as a starting point to comprehend and define quantum entropy. The R\'enyi entropy is given by the equation:
\begin{equation}
Sh_{n}(l)=\frac{1}{1-n}\ln\sum_{\nu=1}p_{\nu}^{n},
\end{equation}
where $p_{\nu}$ represents the associated probabilities and $n\geq0$. When $n=1$, the Shannon entropy is retrieved as $Sh_1=-\sum p_{\nu}\ln p_{\nu}$. These concepts can be employed to explore quantum many-body systems, including quantum phase transitions. For example, consider a quantum system in its ground state $|\psi_0\rangle$. As there is usually an infinite number of observables in quantum mechanics, we can express the ground state as:
\begin{equation}
|\psi_0\rangle=\sum_{\nu}a_{\nu}|\nu\rangle,
\end{equation}
where $|\nu\rangle$ is an eigenstate of the considered observable and $p_{\nu}=|a_{\nu}|^2$ is the corresponding probability of obtaining the eigenvalue. These probabilities can be used to define the Shannon-R\'enyi entropy.

One can use the Shannon-R\'enyi entropy to detect phase transitions by selecting a suitable local observable such as spin~\cite{Stephan2009}. 
At a more fundamental level, it is possible to examine the critical probabilities of a system using a single configuration probability. These probabilities are known as formation probabilities and provide information about the universal boundary entropies. Configurations that exhibit translational invariance at the critical point can evolve into boundary conformal states, such that the subleading term of the free energy of the periodic system with size $L$ is connected to the universal boundary entropy~\cite{Affleck:1991,ARV2020} as
\begin{align}
\ln p_{\nu}=\ln|\langle \nu|\psi_0\rangle|^2=-\Gamma_{\nu} L+2s_{\nu}+\mathcal{O}\left(\frac{1}{L}\right),
\end{align}
where $2s_{\nu}$ is the universal boundary entropy and $\nu$ is the chosen configuration, for instance, all the spins up or down, which is sometimes called the emptiness formation probability. In the case of the periodic critical XY chain, the above formula can be derived analytically for  various configurations in both the $\sigma^x$ and $\sigma^z$ bases~\cite{ARV2020}.

A more intriguing scenario arises when marginal probabilities in a subsystem are considered. These probabilities can be used to determine the central charge and the universality class of critical points~\cite{Stephan_2014,Rajabpour2015,Najafi2016,ARV2020}. In the case where the subsystem is a connected domain with size $l$, the corresponding conformal field theory (CFT) takes the form of a bulk CFT with a slit, subject to a particular conformal boundary condition on the slit. The free energy of such a system is linked to the formation probability as~\cite{Cardy1988,Stephan_2014}:
\begin{align}
\ln p_{\nu}=\ln\frac{Z_{CFT}^{slit}}{Z_{CFT}},
\end{align}
where $Z_{CFT}^{slit}$ ($Z_{CFT}$) is the partition function of CFT in the presence (absence) of the slit.
When the system has periodic boundary conditions there is a leading linear term, along with a subleading logarithmic term that is associated with the central charge of the system. i.e.,~\cite{Stephan_2014}:
\begin{align}
\ln p_{\nu}=-a_{\nu} l-\frac{c}{8}\ln\left[\frac{L}{\pi}\sin\frac{\pi l}{L}\right]+a^0_{\nu}+\mathcal{O}\left(\frac{\ln l}{l}\right).
\end{align}
When $L\to\infty$ the above formula can be derived exactly for the XY chain in both the $\sigma^x$ and $\sigma^z$ bases \cite{ARV2020}. For results regarding the emptiness formation probability in the $\sigma^x$ basis, see~\cite{Franchini2005}

In the case of two disconnected sub-regions, the situation is similar, and we have two slits with conformal boundary conditions. In this context, the free energy of the two slits can be viewed as the Casimir free energy  and the corresponding probabilities also provide information about the critical exponents \cite{Rajabpour2015,Rajabpour2016}. A similar scenario to the one described above arises when we consider all probabilities and compute the R\'enyi entropy. It is expected that the Shannon-R\'enyi entropy for local observables follows a volume-law, where the entropy is proportional to the size of the system, so that the leading term is non-universal and less interesting. However, several numerical studies suggest that the subleading term has a universal property so that we have \cite{Stephan2009}:
\begin{align}
Sh_{n}=\Gamma_{n} L+ \gamma_{n}+\mathcal{O}\left(\frac{1}{L}\right),
\end{align}
where $\gamma_{n}$ is a universal term whose nature is not yet fully understood.

When we take a subsystem instead of the full system and calculate the Shannon-R\'enyi entropy, the subleading term has a logarithmic dependence on the subsystem size, which has a universal coefficient associated with the central charge of critical quantum chains~\cite{Alcaraz_2013,Stephan_2014B,MAR2014}. The general picture is as follows:
\begin{align}
Sh_{n}(l)=a_{n}l+\frac{c_{n}}{8}\ln\left[\frac{L}{\pi}\sin\frac{\pi l}{L}\right]+d_{n}+\mathcal{O}\left(\frac{\ln l}{l}\right).
\end{align}

In general, as $n$ becomes sufficiently large, the $c_{n}$ for specific bases always become dependent on the central charge. However, for smaller values of $n$, the behavior of $c_{n}$ is model dependent, and in some instances it may depend on the central charge, while in others, its nature is not yet well understood. When the system has $U(1)$ symmetry and the central charge is like that of the XXZ chain, one can make plausible arguments based on the Luttinger liquid picture and show the dependency to the central charge for all values of $n$~\cite{Stephan_2014B}; see the main text for the explicit formulas in the $\sigma^{x(z)}$ basis. Nevertheless, in the absence of the $U(1)$ symmetry, even when the central charge is an integer, there isn't a clear dependence on the central charge  for $n\leq 1$~\cite{Tarighi2022}. Note that if one takes a random basis, the coefficient of the logarithm might not have an obvious dependence on the central charge~\cite{MAR2014}. It seems there are particular {\it{conformal}} bases for which the coefficient is directly related to the central charge.

As discussed earlier, the Shannon-R\'enyi entropy in local bases exhibits a volume law trend accompanied by a secondary logarithmic term. This term manifests a sought-after universal behavior that correlates with the central charge. To single out this term, it is advantageous to introduce the following quantity:
\begin{equation} 
I_n(l,L-l)=Sh_n(l)+Sh_n(L-1)-Sh_n(L),
\label{eqn:SI_In}
\end{equation}
which we refer to as R\'enyi difference measure (RDM). This quantity is far more suitable for fitting the data in order to determine the central charge. Although the Shannon-R\'enyi entropy of local observables in a subsystem is a natural quantity from an experimental standpoint, it is theoretically interesting to obtain a ``basis-independent'' quantity by minimizing over all possible bases. This minimization leads to the von Neumann entanglement entropy and its generalizations, including quantum R\'enyi entropies:
\begin{equation}
S_{n}=\frac{1}{1-n}\ln\text{tr} \rho^{n},
\label{SI_quantum Renyi}
\end{equation}
where $\rho$ is the reduced density matrix. For $n=1$, we recover the von Neumann entropy $S=-\text{tr}\rho\ln\rho$. From this perspective the von Neumann entropy is a special case of the Shannon entropy in which one does the measurement in the Schmidt basis.

For quantum chains that demonstrate conformal symmetry, particularly at their critical point and under periodic boundary conditions, the anticipated formula for the von Neumann-R\'enyi entropy is given by:~\cite{Holzhey_1994}:
\begin{equation}
S_n(l) = \frac{c}{6}\frac{n+1}{n}\ln \left(\frac{L}{\pi a}\sin{\frac{\pi l}{L}}\right)+\mathcal{O}(1), 
\label{eqn:SI_CFT-EE-PBC}
\end{equation}
where $c$ is the central charge and $a$ is the lattice constant.

So far, we have demonstrated that the logarithmic coefficient in both the von Neumann-R\'enyi entanglement entropy and the Shannon-R\'enyi entropy is reliant on the central charge under periodic boundary conditions. This relationship also holds for open boundary conditions, provided the subsystem is selected from one end of the chain; here, the logarithmic coefficient is merely half of that in the periodic case. Table~\ref{tab:I1} consolidates all these equations, encompassing the scenario where the system is infinite and the subsystem is centered.

\begin{table}[ht]
\centering 
\begin{tabular}{|c|c|c|}
\hline\hline 
\textbf{Boundary condition} & \textbf{$S_n$}  & \textbf{$I_n$} \\ [0.5ex]
\hline 
Infinite  & $\frac{c}{6}\frac{n+1}{n}\ln (l/a)$ & $\frac{c_n}{4}\ln (l/a)$ \\ [0.75ex]  
\hline
Open & $\frac{c}{12}\frac{n+1}{n}\ln (\frac{2L}{\pi a}\sin{\frac{\pi l}{L}})$ & $\frac{c_n}{8}\ln(\frac{2L}{\pi a}\sin{\frac{\pi l}{L}}) $ \\ [0.75ex]
\hline
Periodic & $\frac{c}{6}\frac{n+1}{n}\ln (\frac{L}{\pi a}\sin{\frac{\pi l}{L}})$ & $\frac{c_n}{4}\ln(\frac{L}{\pi a}\sin{\frac{\pi l}{L}})$ \\ [0.75ex] 
\hline 
\end{tabular}
\caption{Leading scaling behavior of the von Neumann-R\'enyi entropy $S_n$, and the quantity $I_n$ related to the Shannon-R\'enyi entropy, for different boundary conditions with respect to the subsystem size.} 
\label{tab:I1}
\end{table}

Based on the boundary conditions, whether open or periodic, and the specific measure, whether von Neumann-R\'enyi entanglement entropy or Shannon-R\'enyi entropy (in different bases) for varying values of $n$, there exists an abundance of methods to ascertain the central charge. Each of these measures exhibits a different degree of finite size scaling (FSE) in such the effect being significantly more pronounced than others. Generally, the Shannon-R\'enyi entropy under open BC tends to exhibit the most substantial FSE independent of the choice of the bases, followed by the von Neumann-R\'enyi entanglement entropy with open BC, and this tendency is usually consistent across different values of $n$. In contrast, periodic BC typically shows the least FSE. Nonetheless, the model and the choice of $n$ can influence whether the von Neumann-R\'enyi entanglement entropy or the Shannon-R\'enyi entropy in particular basis is less affected by FSE. In our approach to determining the central charge, we initially evaluated all these different quantities using numerical calculations at the critical point.
Upon identifying the quantity with minimal FSE, we then used that metric to calculate our central charge. It's important to note that the von Neumann-R\'enyi entanglement entropy is only accessible when the Hamiltonian is stoquastic in its natural basis, which allows for reconstruction of the ground state. This scenario applies to our work with the transverse field Ising chain. Given this, the Shannon-R\'enyi entropy emerges as a more universally applicable quantity for extracting the central charge. In summary, to obtain the most accurate estimate of the central charge, one should typically employ periodic BC and determine the optimal basis and value of $n$ to ensure the FSE is at its lowest impact.

\subsection{Models and connection to previous works}
\label{SI:models}
In this section, we summarize known facts about the Shannon-R\'enyi entropy in the TFI chain and the XXZ chain.
We start with the transverse-field Ising Hamiltonian
\begin{eqnarray}\label{eq:SI-XY_hamiltonian}\
\textbf{H}_{1}=-J\sum_{l=1}^{L}\sigma_l^z\sigma_{l+1}^z-h\sum_{l=1}^L\sigma_l^x\;,
\end{eqnarray}
where $\sigma^{\alpha}$ denotes a Pauli operator, $J$ is the coupling strength between neighboring spins, and $h$ corresponds to the external transverse magnetic field. For $h/J=1$, the ground state of the system is critical and it can be described by a CFT with the central charge $c=\frac{1}{2}$. The $h/J=1$ point separates the disordered phase $h/J>1$ from the ordered phase $h/J<1$ corresponding to spontaneous breaking of the $\mathbb{Z}_2$ symmetry. Interestingly, this Hamiltonian is part of a category of Hamiltonians known as stoquastic Hamiltonians~\cite{bravyi2007} . Essentially, when the ground state (GS) is expressed in the $\sigma^z$ or $\sigma^x$ basis, all amplitudes are non-negative real numbers. This indicates that by performing a projective measurement in one of these specific bases, the state can be unambiguously reconstructed: $|\psi_0\rangle=\sum_{\nu}\sqrt{p_{\nu}}|\nu\rangle$. Once the state is reconstructed, various desired quantities, including the von Neumann-R\'enyi entanglement entropy, can be computed. On the other hand, these probabilities can be also  employed directly to determine the Shannon-R\'enyi entropies introduced in Eq.~\ref{eqn:Shannon_L}. At the critical point, it is anticipated that $c_n=\frac{n}{2(n-1)}$ will hold for $n>1$ in the presence of periodic boundary conditions~\cite{Stephan_2014B,MAR2014}. For open boundary conditions, when the subsystem originates from one end of the spin chain, the anticipated coefficient is half of this value, though it is accompanied by significantly pronounced finite size effects~\cite{Stephan_2014B}.

Another quantum spin chain that has been widely studied is the XXZ chain, which is defined by
\begin{eqnarray}
\textbf{H}_{2}=-\sum_{l=1}^{L-1}\Big{(}\sigma_l^x\sigma_{l+1}^x+\sigma_l^y\sigma_{l+1}^y+\Delta\sigma_l^z\sigma_{l+1}^z\Big{)}\;, 
\label{eqn:SI_xxz}
\end{eqnarray}
where, $\Delta$ is the anisotropy parameter. For $-1\leq \Delta <1$, the system is critical and gapless and its ground state is described by the  compactified boson CFT with a central charge of $c=1$ \cite{Giamarchi}. This allows us to delve into the universality concerning the variations in the anisotropy parameter $\Delta$. Specifically, for $\Delta<0$, the dominant probability in the $\sigma^z$ basis exhibits antiferromagnetic traits. These are believed to undergo renormalization to Dirichlet boundary conditions within the Luttinger liquid depiction of the XXZ chain, suggesting  links with the inherent CFT~\cite{Stephan_2014B,MAR2014}. In this scenario, the coefficient of the logarithm, $c_n$, present in equation~\eqref{eqn:mutual_l}, is influenced by both $\Delta$ and $n$. It's postulated that a discontinuity emerges at $n^{z}_t(n^{x}_t)=\frac{2\pi}{\arccos \Delta}(\frac{2}{\pi}\arccos \Delta)$~ in the $\sigma^{z}(\sigma^{x})$ basis so that we have $c_n=1$ for $n<n_t$ and $c_n=\frac{n}{n-1}$ for $n>n_t$. The ground state amplitudes of the XXZ chain in both the $\sigma^x$ and $\sigma^z$ bases aren't solely non-negative values, preventing the state's reconstruction using just the configuration probabilities from one of these bases. For this particular model, we exclusively explore the Shannon-R\'enyi entropies. It's important to mention that near the transition point $n=n_t$, the finite size effects are significantly pronounced, which makes the precise computation of $c_n$  challenging.

\begin{figure*}[hbt!]
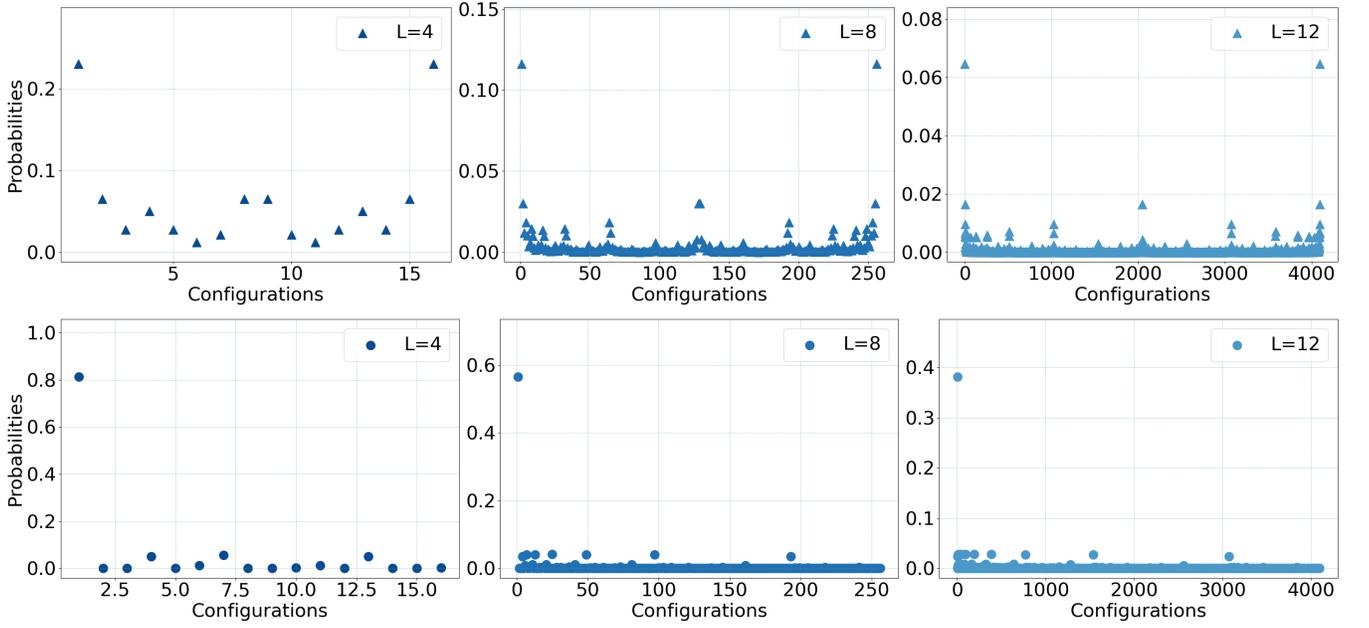

    \centering
    \includegraphics[width=\linewidth]{figures/supplementary/Prob_TFIM_xyz_op_minor.pdf}
    \includegraphics[width=\linewidth]{figures/supplementary/Prob_TFIM_zyx_op_minor.pdf}
    \caption{Distribution of configuration probabilities for transverse field Ising model at critical point ($h/j=1$) with open boundary condition in the $\sigma^z$ and $\sigma^x$ bases shown in top and bottom row respectively. In the $\sigma^z$ basis, due to $\mathbb{Z}_2$ symmetry, flipping the spins will leave the probabilities invariant manifesting in mirrored probabilities. In the $\sigma^x$ basis, configurations with odd magnetization are not allowed leading to the absence of those configurations.}
    \label{fig:Prob_TFIM_open}
\end{figure*}

\subsection{Symmetry and basis choice}
\label{SI:basis choice}

\begin{figure*}[hbt!]
    \centering
    \includegraphics[width=\linewidth]{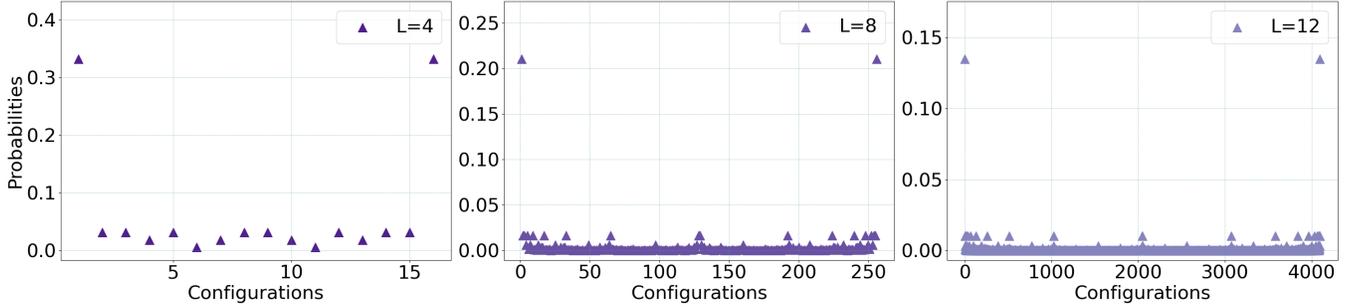}
    \caption{Distribution of configuration probabilities for the transverse field Ising model at critical point ($h/j=1$) with periodic boundary condition in the $\sigma^z$ and $\sigma^x$ bases shown in the top and bottom row, respectively. In the $\sigma^z$ basis, we observe that $\mathbb{Z}_2$ symmetry manifests itself in probabilities. In the $\sigma^x$ basis, configurations with odd magnetization are not allowed and thus do not appear.}
    \label{fig:Prob_TFIM_periodic}
\end{figure*}

The amplitude and corresponding probabilities will depend on the measurement basis. Here, we present the results for two choices of basis of the TFI and XXZ models with both open and periodic boundary conditions.

We start with the transverse Ising model with the open boundary condition. The amplitudes of the ground state in the $\sigma^z$ basis are real and non-negative, meaning that one can reconstruct the quantum state from the measured probabilities. Also, due to the $\mathbb{Z}_2$ symmetry, the configuration probabilities are invariant with respect to flipping spins or exchanging up with down. Thus in the binary representation, the second half of the probabilities are mirroring the first half, as seen in Fig.~\ref{fig:Prob_TFIM_open}. On the other hand, in the $\sigma^x$ basis, due to $\mathbb{Z}_2$ parity number symmetry, the configurations with odd parity are not allowed. In Fig.~\ref{fig:Prob_TFIM_open}, we have depicted the corresponding probabilities for various system size of $L=4,8,12$
clearly indicating that many of the configurations are absent.  In the case of periodic boundary conditions in the $\sigma^z$ basis, the symmetries are even more pronounced; and on top of reflection symmetry, we observe that many of the probabilities are identical.  Similarly in the $\sigma^x$ basis, many configurations are not allowed and lead to zero probabilities for those configurations see Fig.~\ref{fig:Prob_TFIM_periodic}. 

\begin{figure*}[hbt!]
    \centering
    \includegraphics[width=\linewidth]{figures/supplementary/Prob_XXZ_xyz_op_minor.pdf}
    \includegraphics[width=\linewidth]{figures/supplementary/Prob_XXZ_zyx_op_minor.pdf}
    \caption{ Distribution of configuration probabilities for the XXZ model at critical point ($\Delta=-0.5$) with open boundary condition in the $\sigma^z$ and $\sigma^x$ bases, shown in the top and bottom row, respectively. }
     \label{fig:Prob_XXZ_open}   
\end{figure*}
\begin{figure*}[hbt!]
    \centering
    \includegraphics[width=\linewidth]{figures/supplementary/Prob_XXZ_xyz_per_minor.pdf}
    \includegraphics[width=\linewidth]{figures/supplementary/Prob_XXZ_zyx_per_minor.pdf}
    \caption{Distribution of configuration probabilities for the XXZ model at critical point ($\Delta=-0.5$)with periodic boundary condition in $\sigma^z$ and $\sigma^x$ bases, shown in the top and bottom row respectively.}
    \label{fig:Prob_XXZ_periodic}     
\end{figure*}
In the case of XXZ Hamiltonian, due to $U(1)$ symmetry in the $\sigma^z$-basis, only configurations with half-filling  have non-zero probabilities, see Fig.~\ref{fig:Prob_XXZ_open} and Fig.~\ref{fig:Prob_XXZ_periodic}. 

\subsection{Direct state preparation}
\label{SI:direct_state_prep}

\begin{figure*}[ht!]
    \centering
    \includegraphics[scale=0.3]{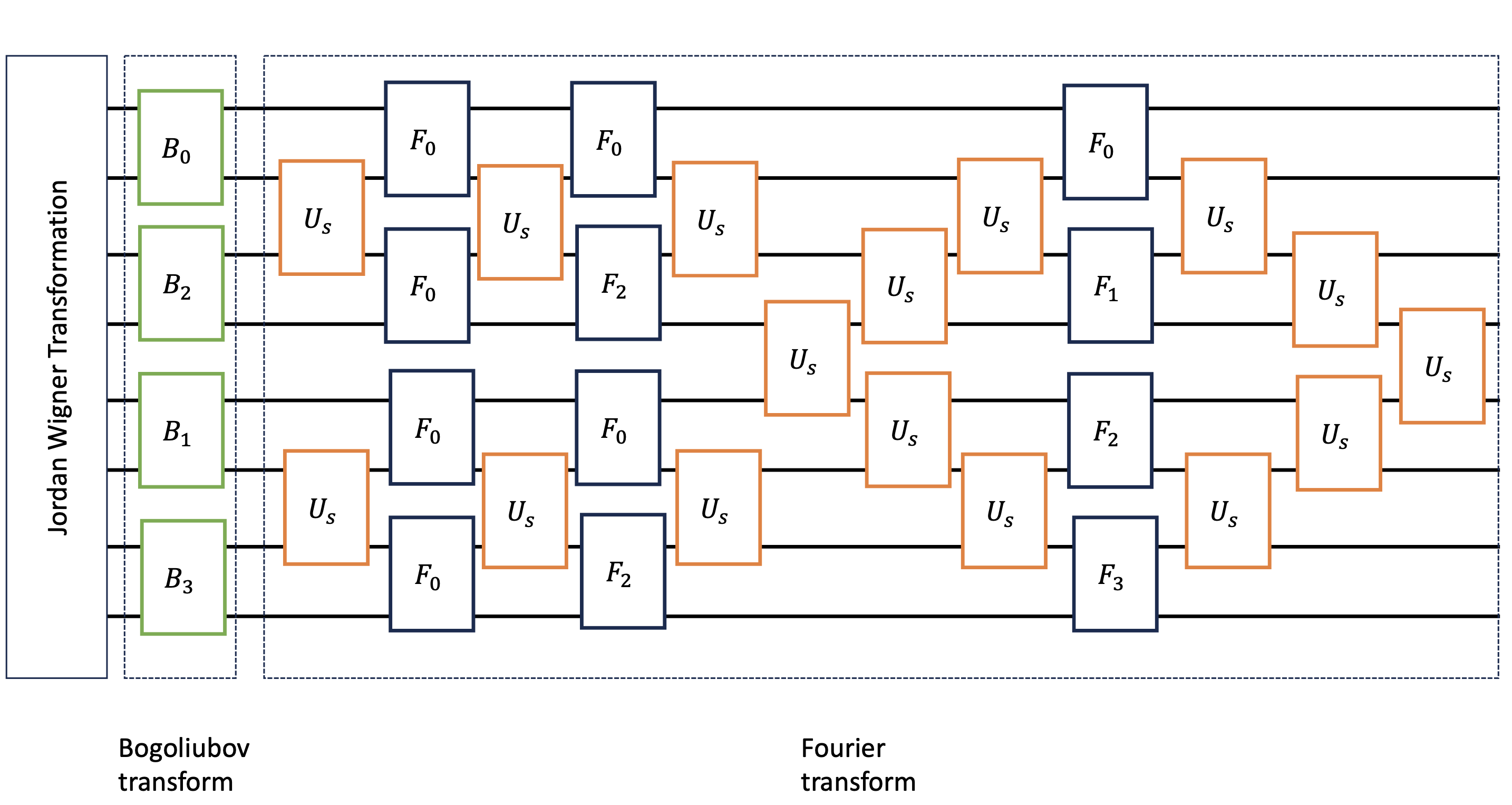}
\caption{A template circuit \cite{verstraete2009quantum} for preparing the ground state of XY Hamiltonian. There are three steps. First, a Jordan-Wigner transformation transforms the spin operators into fermionic modes. Then, we have the Bogoliubov transformation ($B_{i}$ gates). The final step is the Fourier transform ($U_s$ for fermionic swap and $F_{i}$ for fast Fourier transform).}
    \label{fig:direct state prep}
\end{figure*}

In this work, we have used variational quantum algorithms (VQAs) to prepare the ground state of various interacting spin systems. VQAs allow us to prepare the ground state with relatively short-depth hardware-efficient circuits. However, in order to achieve high fidelity in preparing the ground state, these hybrid approaches sometimes require prohibitively many iterations between the quantum and classical computers. Alternatively, for a set of integrable models, it is possible to find a unitary transformation that simplifies the Hamiltonian into a non-interacting system. In this section we summarize the method from Ref.~\cite{verstraete2009quantum}. Thus, in this section we investigate the possibility of implementing such protocols.  In Fig.~\ref{fig:direct state prep}, we provide a template circuit of this transformation. There are three types of gate involved. First we have the Bogoliubov transformation gate, parameters for which depend on the external magnetic field and the anisotropy parameter of the Hamiltonian under study. Second, to implement the Fourier transform, we use a fermionic swap operation for implementing the anti-commuting properties of the fermions and a Fourier transform gate for implementing the relative phase change in fast Fourier transform. 

Starting with the XY Hamiltonian with anisotropic parameter $\gamma$ and external field $\lambda$: 
\begin{align}
H_{XY} &= \sum_{i=1}^{L} \left( \frac{1 + \gamma}{2} \sigma^x_i \sigma^x_{i+1} \right. \nonumber 
\quad \left. + \frac{1 - \gamma}{2} \sigma^y_i \sigma^y_{i+1} \right) \nonumber 
+ \lambda \sum_{i=1}^{L} \sigma^z_i \nonumber \\
&+ \frac{1 + \gamma}{2} \sigma^y_1 \sigma^z_2 \dots \sigma^z_{L-1} \sigma^y_L\nonumber 
+ \frac{1 - \gamma}{2} \sigma^x_1 \sigma^z_2 \dots \sigma^z_{L-1} \sigma^x_L,
\end{align}
the Jordan-Wigner transformation leads to the fermionic modes resulting in the following Hamiltonian: 
\begin{align}
H_2[c] &= \frac{1}{2} \sum_{i=1}^{L} \left( c^\dagger_{i+1} c_i + c^\dagger_i c_{i+1} \right) + \\
&\gamma \left( c^\dagger_i c^\dagger_{i+1} + c_i c_{i+1} \right) 
+\lambda \sum_{i=1}^{L} c^\dagger_i c_i.
\end{align}
Consequently, we apply the Fourier transform on the fermionic model:
\[
b_k = \frac{1}{\sqrt{L}} \sum_{j=1}^{L} e^{i \frac{2\pi k j}{L}} c_j \; , \; k = -\frac{L}{2}+1,...,\frac{L}{2}.
\]
In Fig.~\ref{fig:direct state prep}, $U_s$ are the quantum gate for fermionic SWAP operations 
$$U_{SWAP} = \begin{pmatrix}
1 & 0 & 0 & 0 \\
0 & 1 & 0 & 0 \\
0 & 0 & 1 & 0 \\
0 & 0 & 0 & -1
\end{pmatrix}.$$
Thus we use the $F_{k}$ gates to change the relative phase for Fourier transform with $\alpha(k)=exp(i2\pi k/L)$
$$F_{k} = \begin{pmatrix}
1 & 0 & 0 & 0 \\
0 & 1/\sqrt{2} & \frac{\alpha(k)}{\sqrt{2}} & 0 \\
0 & 1/\sqrt{2} & -\frac{\alpha(k)}{\sqrt{2}} & 0 \\
0 & 0 & 0 & -\alpha(k)
\end{pmatrix}.$$

We then perform a Bogoliubov transform leading to a full disentanglement of the original Hamiltonian with the momentum dependent mixture of modes 
$$
a_{k} = \cos{(\theta_{2}/k)}b_{k} - i\sin{(\theta_{k}/2)}b^{\dagger}_{-k},
$$
where
$$\theta_k = \arccos \left( \frac{-\lambda + \cos \left( \frac{2\pi k}{L} \right)}{\sqrt{\left( \lambda - \cos \left( \frac{2\pi k}{L} \right) \right)^2 + \gamma^2 \sin^2 \left( \frac{2\pi k}{L} \right)}} \right).
$$
With this, the full Hamiltonian can be written as 
$$H_4[a] = \sum_{k=-\frac{n}{2}+1}^{\frac{L}{2}} \omega_k a^\dagger_k a_k,
$$
with
$$\omega_k = \sqrt{ \left( \lambda - \cos \left( \frac{2 \pi k}{L} \right) \right)^2 + \gamma^2 \sin^2 \left( \frac{2 \pi k}{L} \right) }.$$
Finally, we perform this disentanglement using the $B_k$ gate
$$
B_k = \begin{pmatrix}
\cos \theta_k & 0 & 0 & i \sin \theta_k \\
0 & 1 & 0 & 0 \\
0 & 0 & 1 & 0 \\
i \sin \theta_k & 0 & 0 & \cos \theta_k
\end{pmatrix}.
$$
 \begin{figure*}[hbt!]
    \centering
    \includegraphics[scale=0.3]{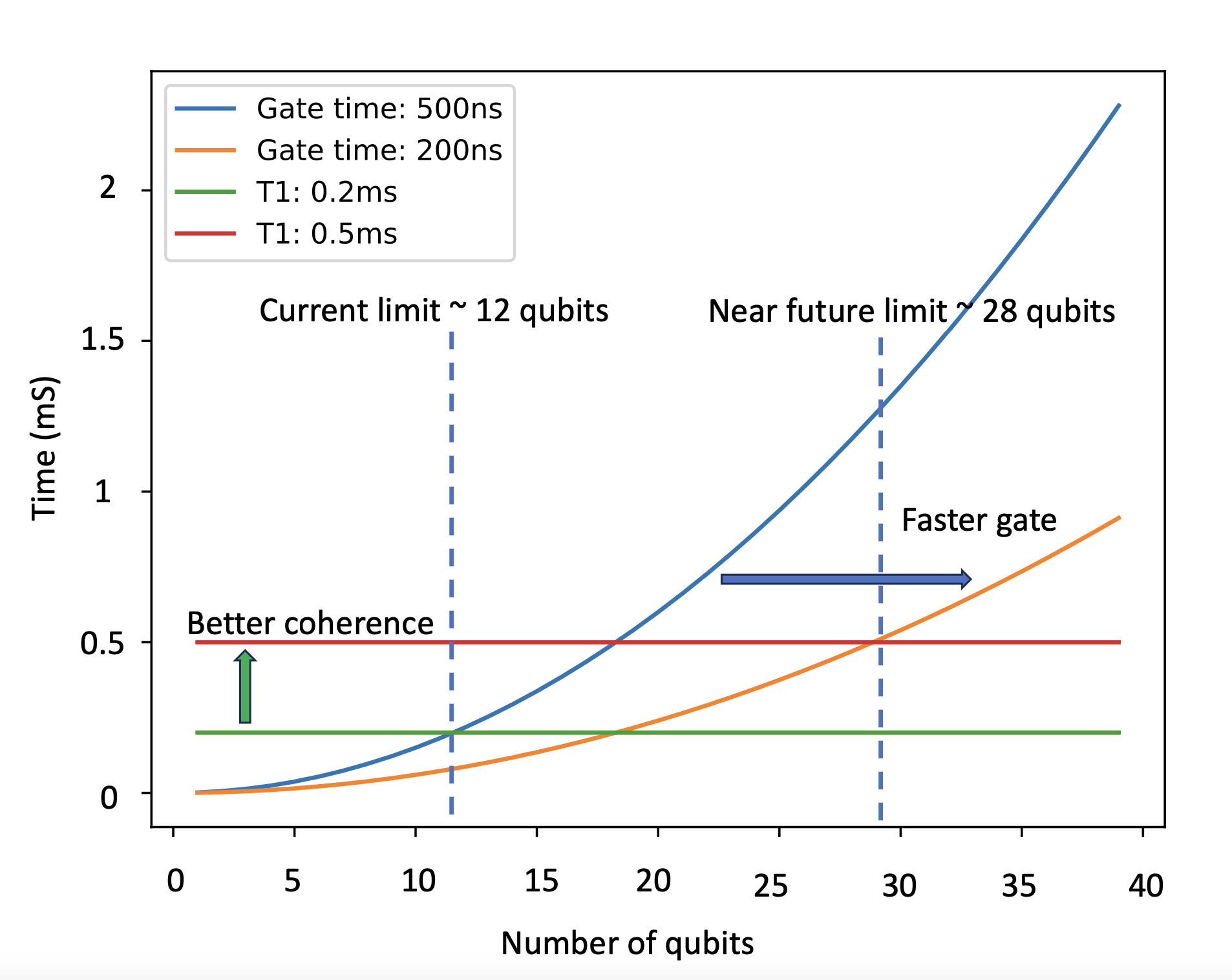}
\caption{Cost of directly preparing the ground state of the XY models:
We present a plot depicting the total evolution time required (shown in blue) as a function of the number of qubits for the direct preparation of ground states in XY models. Based on the circuit depth, current coherence times and gate times of the hardware, we are limited to about 12 qubits. But modest improvements in these number (as seen in IBM's next generation hardware) can extend the scope to about 28 qubits.}
    \label{fig:cost}
\end{figure*}

Following the exact state preparation methods described in \cite{verstraete2009quantum} for XY models, a system with $L$ spins will require a total of $L^2$ two-qubit gates. Since we need 3 CNOT gates to construct each two-qubit gate, the total CNOT count is $3L^2$. We compared the cost of this direct state preparation method with the specifications of the current hardware in Fig.~\ref{fig:cost}. We calculated the total time required to prepare the ground state for up to 127 qubits and compared it with the current coherence times. Furthermore, we made rough estimates of the circuit fidelity based on pessimistic and optimistic qualities of the entangling operations. Overall, we can expect to apply this direct state preparation technique for up to approximately 40 qubits in the near term. With improvements in coherence times, we can expect to push the system size to around 60 qubits. Our numerical simulations demonstrate that system sizes like these are sufficient for accurate estimation of the central charge for various spin systems. Therefore, direct state preparation offers a promising approach to explore universality classes using near-term digital quantum computers.

\subsection{State preparation with the Krylov quantum diagonalization algorithm}
\label{SI:Krylov method}

 Subspace methods use a set of states as a basis for the construction of a smaller representation of a matrix of interest $\hat{H}$ (the Hamiltonian in our case), which captures approximations to its essential properties (e.g., lowest eigenvalue). One possibility for such subspaces is the one constructed by a reference state $\vert \psi \rangle$ and states obtained by applying successive powers of the matrix $\hat{H}$ up to $r-1$. This is called the power Krylov subspace $K^{(r)}$ of order $r$:

\begin{equation}
    K^{(r)} = \left\{ \vert \psi \rangle, \hat{H} \vert \psi \rangle, \hat{H}^2 \vert \psi \rangle, ..., \hat{H}^{r-1} \vert \psi \rangle \right\} .
\end{equation}

\noindent
Lanczos algorithm is a classical algorithm that uses a variant of the power Krylov subspace (where the basis states are made orthonormal), to provide a compressed and faithful representation of the original matrix (which should be sparse and Hermitian). It is widely used for studying the time evolution and the ground state of large quantum systems. However, for entangled many-body states the method runs into an exponential increase in memory requirements for storing the state vector. A quantum version of the method \cite{motta2020determining, cortes2022quantum} avoids this issue because of the polynomial footprint of storing large entangled states on a quantum memory. In this case, the impossibility of implementing $\hat{H}$ as an operation on a quantum computer (a unitary operator is needed) leads to the use of the time-evolution operator $\hat{U}=e^{-i\hat{H}t}$ generated by $\hat{H}$. Therefore we can prepare states of what has been termed \cite{parrish_2019, stair2020krylov,urbanek2020chemistry,cohn2021filterdiagonalization,klymko2022realtime,jamet2022greens,tkachenko2024davidson,lee2023sampling,kirby2023exactefficient,shen2023realtimekrylov,motta2020determining,seki2021powermethod,cortes2022quantum} the unitary Krylov subspace $K_U$ on a quantum computer

\begin{equation}
    K_U^{(r)} = \left\{ \vert \psi \rangle, \hat{U} \vert \psi \rangle, \hat{U}^2 \vert \psi \rangle, ..., \hat{U}^{r-1} \vert \psi \rangle \right\} ,
\end{equation}

\noindent
where different powers of $\hat{U}$ correspond to time-evolvution for different time steps. We can then project the Hamiltonian onto the subspace to construct a compact representation $T_{jl} = \langle \psi_j \vert \hat{H} \vert \psi_l \rangle$. Once the matrix elements are calculated on a quantum computer, the matrix $T$ can be stored and processed classically. Since the basis states of the Krylov subspace are not orthogonal $\langle \psi_j \vert \psi_l \rangle \neq 0$ we cannot directly diagonalize the matrix $T$ as $T c = t c$ where $t$ are the eigenvalues of $T$ and $c$ its eigenvectors. Instead, we need to take into account the metric of the space, which means solving the generalized eigenvalue problem 

\begin{equation}
    K^t \hat{H} K c = t K^t K c 
\end{equation}

\noindent
where $T = K^t \hat{H} K$ and we have defined $S = K^t K$. The matrix elements of $T$ and $S$ can both be obtained using a quantum computer. From the solution of the generalized eigenvalue problem one can then get the lowest eigenvalue $t_{min}$ corresponding to the ground state energy estimate. Furthermore, the corresponding eigenstate $c_{min}$ is an estimate of the ground state and its coefficients tell us the contribution of each state of the Krylov subspace. We could then measure any observable of interest on the states of the Krylov subspace and their expectation values on the ground state can be calculated by weighing the expectation values obtained from the different Krylov subspace basis states by the coefficients found in $c_{min}$.

\begin{figure}[]
    \centering
    \includegraphics[width=\columnwidth]{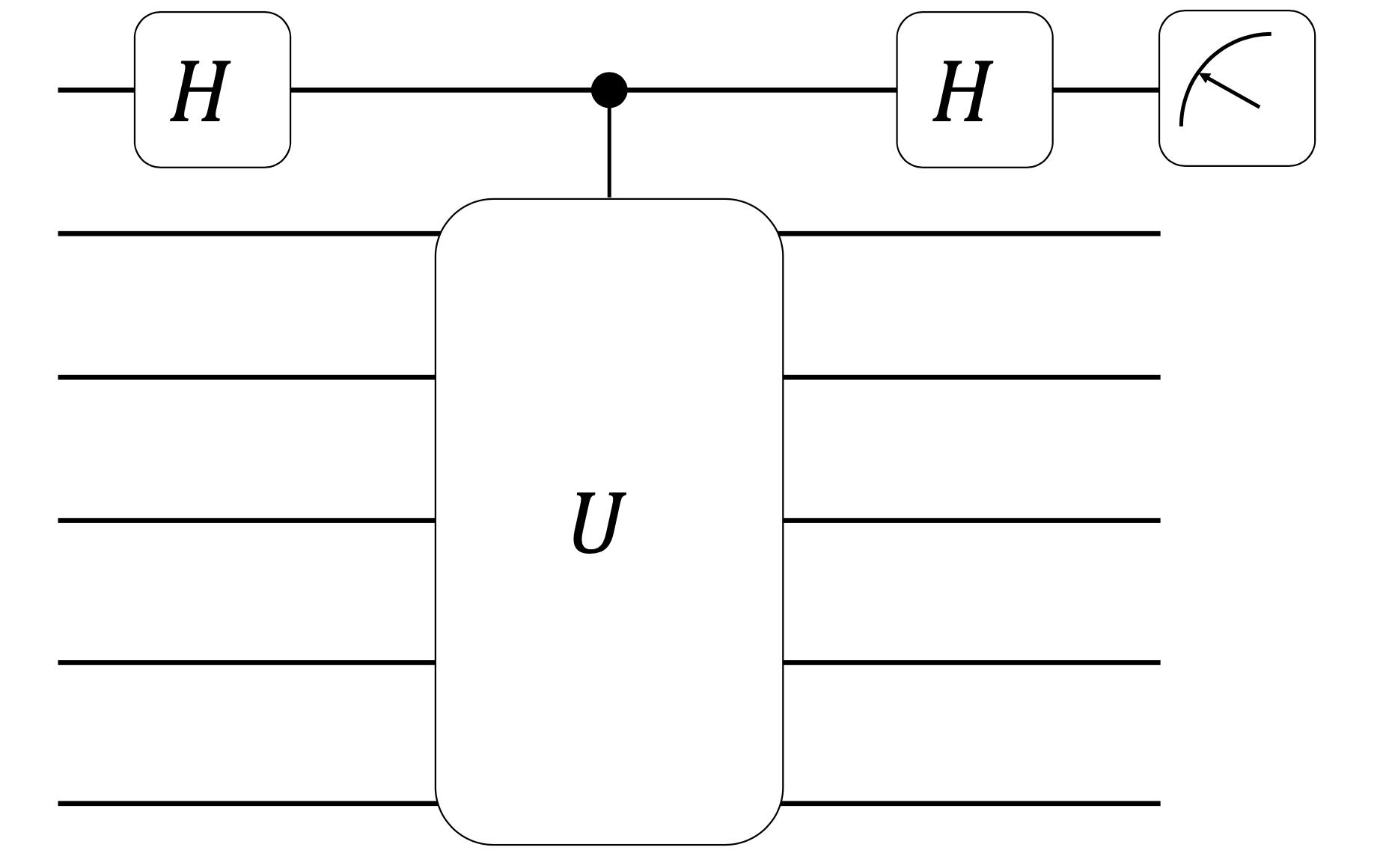} \\
    \includegraphics[width=\columnwidth]{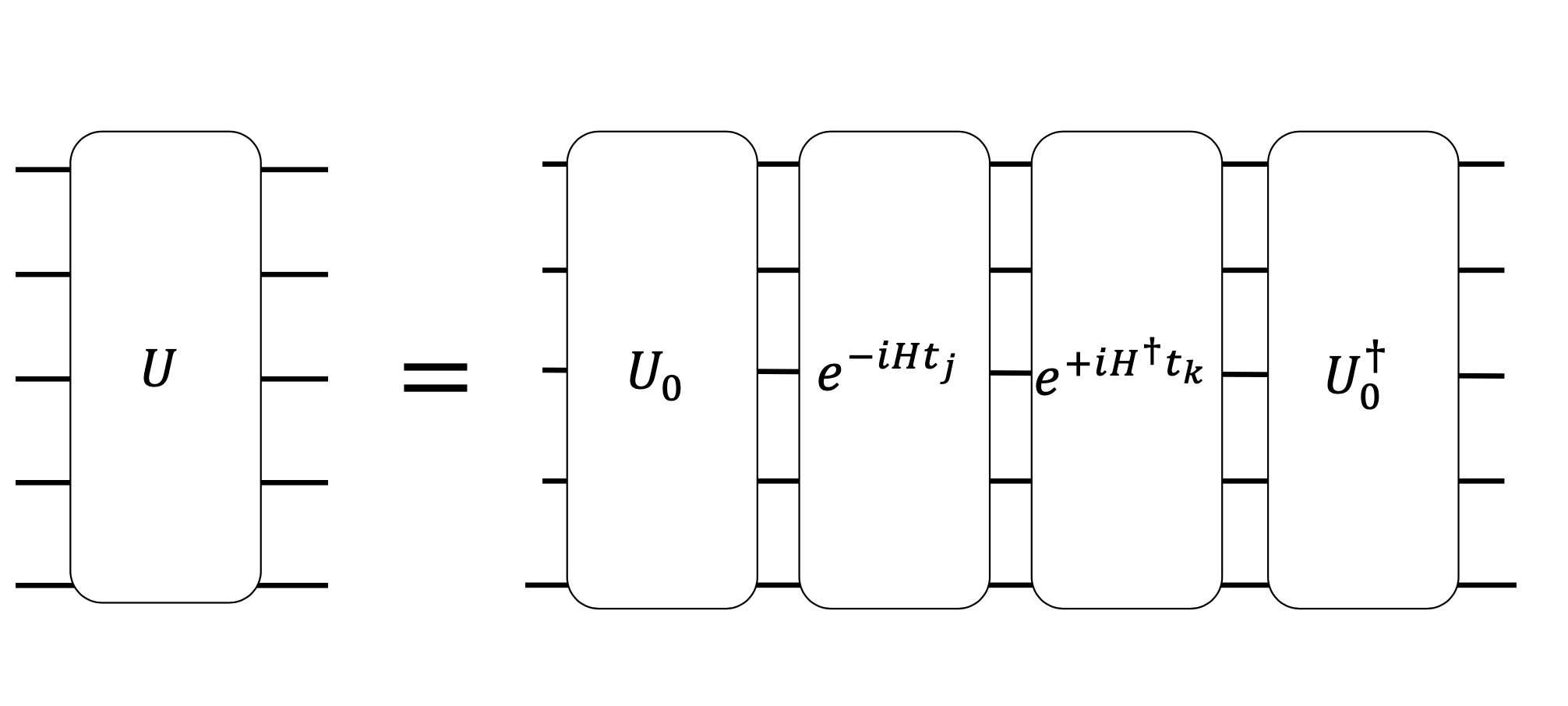}
    \caption{  Example circuit for the calculation of overlaps $S_{jk}$ with the Hadamard test. Top panel: Hadamard test circuit for the real part of a matrix element of $S$. The different time steps in the controlled unitary $U$ will yield different matrix elements of $S$. Bottom panel: decomposition of the controlled unitary $U$ in its component unitaries. $U_0$ prepares the reference state $U_0 \vert 0 \rangle = \vert \psi \rangle$ and the time-evolution operator $e^{-iHt_k}$ gives the time-evolved reference state $\vert \psi_k \rangle =e^{-iHt_k}\vert \psi \rangle $  }
     \label{fig:had_test}   
\end{figure}

\begin{figure}[]
    \centering
    \includegraphics[width=\columnwidth]{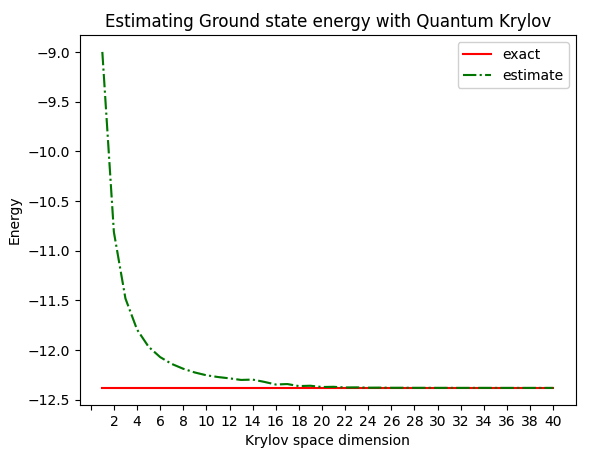}
    \caption{Estimated ground state energy with the Krylov quantum diagonalization method for the transverse field Ising Hamiltonian $H=J \sum_{i,j}^L \sigma^z_i \sigma^z_j + h \sum_i^L \sigma^x_i$ with $J=-1$ and $h=-1$ on ten spins with open boundary conditions.}
     \label{fig:krylov_gs}   
\end{figure}

Fig.~\ref{fig:had_test} shows a simple example of a quantum circuit which can be used to calculate the matrix elements of $S$ using the Hadamard test. A similar circuit, with the extra controlled application of the Hamiltonian terms can be used to measure the matrix elements $T_{jl}$. In this scenario, we can give an estimate for the total number of circuits and the depth of the uncontrolled unitary $U$ of the quantum circuits. Accounting for the symmetry of $T$ and $S$, we only need to measure $r$ different matrix elements (where $r$ is the dimension of the Krylov subspace). We then need to carry out two separate Hadamard tests to measure the real and imaginary part of each matrix element. This means that we have a total of $4r$ different circuits to determine all the matrix elements of $T$ and $S$. Each circuit consists of a controlled state preparation and a Trotter evolution for both states that we are calculating the overlap of. Thus, the total depth (before any optimization) of the uncontrolled unitary is twice the sum of the state preparation and Trotter circuit depths. Once $U$ is turned into a controlled unitary, the circuit depth increases greatly. Especially considering the additional SWAP operations needed to route the ancilla qubit and control the operations on all the other qubits. In theory, one could take advantage of symmetries in the Hamiltonian of interest, like $U(1)$ symmetry to optimize the Hadamard test circuit to obtain a much more efficient implementation\cite{cortes2022quantum,yoshioka2024diagonalization}, however the implementation becomes nontrivial in which we postpone it for future studies.  As an example, we show the convergence of the estimated ground state energy of the transverse field Ising Hamiltonian for chain of ten spins in Fig.~\ref{fig:krylov_gs}. 

\section{Experimental workflow}\label{sec:implementation}
\label{SI:experimental}

In this section, we delve deeper into the experimental procedures and data processing techniques. We begin with a detailed explanation of the ground state preparation method employed in our experiment. Following this, we offer a comprehensive overview of probabilistic error cancellation, which serves as our primary error mitigation technique. Finally, we present an in-depth discussion of the data post-processing and error analysis methods used to extract the central charge values.

\subsection{Experimental set-up and choice of qubits}
\label{SI:qubit_choice}

We used IBM's 27 qubit Falcon and 65 qubit Hummingbird processors. These processors are organized in a heavy-hex topology and two qubit entangling operations are performed using cross-resonance interaction of fixed-frequency transmon qubits. The Hummingbird processor used has a median CX error rate of $9.888 \cdot 10^{-3}$, T1 and T2 times of 168 $\mu s$ and 166 $\mu s$, single qubit error rate of $3.260\cdot 10^{-4}$ and readout error of 
$2.490 \cdot 10^{-2}$. The Falcon processor we used has median CX error rate of $9.448\cdot 10^{-3}$, T1 and T2 times of 117 $\mu s$ and 84 $\mu s$, single qubit error rate of $2.468\cdot 10^{-4}$ and readout error of 
$1.830\cdot 10^{-2}$. The median CNOT gate time was 551 $ns$ for the Hummingbird processor and 450 $ns$ for the Falcom processor. 

\subsection{Simulated variational quantum algorithms for ground state preparation}
\label{SI:VQE}

\begin{figure*}[t!]
    \centering
    \begin{adjustbox}{width=0.8\textwidth}
    \begin{quantikz}
            \lstick{$\ket{+}$} & \gate[6]{U^{(0)}}  & \qw & \ldots \ \ \ \ & \gate{R_y(\theta_{d,0,0})}\gategroup[6,steps=5,style={dashed,rounded
    corners,fill=blue!20, inner
    xsep=2pt},background,label style={label
    position=below,anchor=north,yshift=-0.2cm}]{$U^{(d)}$}
    & \ctrl{1} & \gate{R_y(\theta^{(d)}_{0,1})} & \qw &  \ctrl{5} & \qw & \ldots \ \ \ \  &  \gate{R_y(\theta_{d,0,2})}  & \meter{} \\
            \lstick{$\ket{+}$} && \qw & \ldots \ \ \ \ & \gate{R_y(\theta^{(d)}_{1,0})} & \phase{} & \gate{R_y(\theta^{(d)}_{1,1})} & \ctrl{1} &  \qw  & \qw & \ldots \ \ \ \ & \gate{R_y(\theta^{(d)}_{1,2})}  & \meter{} \\ 
            \lstick{$\ket{+}$} && \qw & \ldots \ \ \ \ & \gate{R_y(\theta^{(d)}_{2,0})} & \ctrl{1} & \gate{R_y(\theta^{(d)}_{2,1})} & \phase{} & \qw  & \qw & \ldots \ \ \ \ &  \gate{R_y(\theta^{(d)}_{2,2})}  & \meter{} \\
            \lstick{$\ket{+}$} && \qw & \ldots \ \ \ \ & \gate{R_y(\theta^{(d)}_{3,0})} & \phase{} & \gate{R_y(\theta^{(d)}_{3,1})} & \ctrl{1} &  \qw & \qw & \ldots \ \ \ \ &  \gate{R_y(\theta^{(d)}_{3,2})} & \meter{} \\ 
            \lstick{$\ket{+}$}  && \qw & \ldots \ \ \ \ & \gate{R_y(\theta^{(d)}_{4,0})} & \ctrl{1} & \gate{R_y(\theta^{(d)}_{4,1})} & \phase{} & \qw  & \qw & \ldots \ \ \ \ &  \gate{R_y(\theta^{(d)}_{4,2})}  & \meter{} \\
            \lstick{$\ket{+}$}  && \qw & \ldots \ \ \ \ & \gate{R_y(\theta^{(d)}_{5,0})} & \phase{} & \gate{R_y(\theta^{(d)}_{5,1})} & \qw & \phase{} & \qw & \ldots \ \ \ \ &  \gate{R_y(\theta^{(d)}_{5,2})} & \meter{}  
\end{quantikz}
\end{adjustbox}
\includegraphics[width=\linewidth]{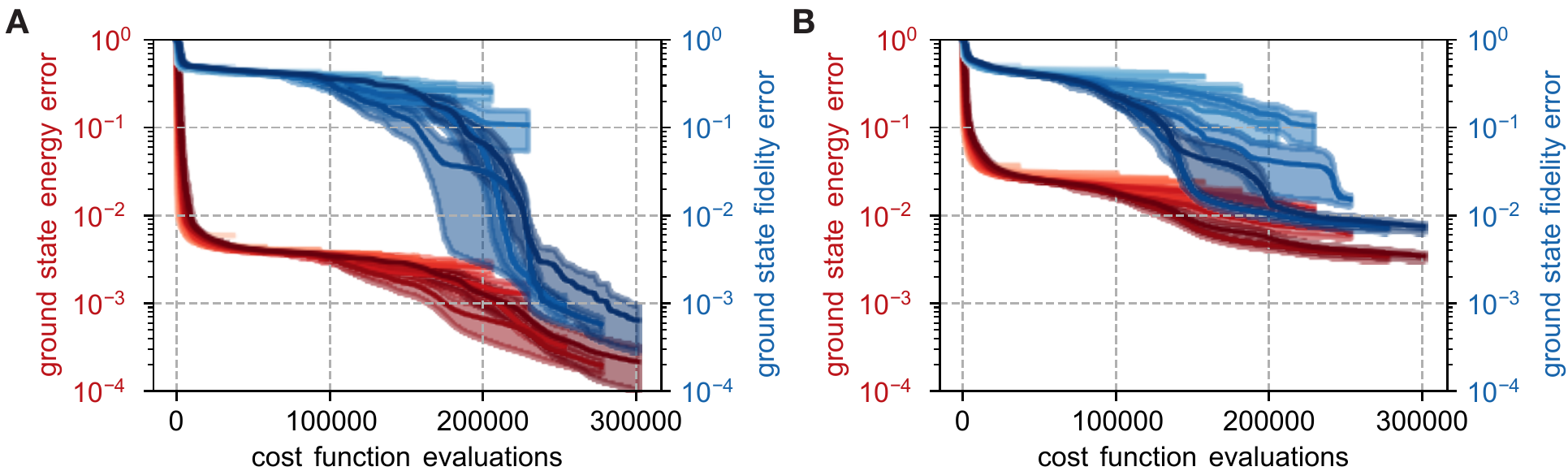}
    \caption{(Top) Initial state preparation (Hadamard $H$ gate on each qubit) followed by layers $U^{(d)}$ of the checkerboard ansatz depicted  for $L=6$ qubits, consisting of parameterized single-qubit $R_Y(\theta)$ gates on each qubit interlaid between parameterized two-qubit $CZ$ gates on alternating even/odd pairs of neighboring qubits. The two-qubit gates do (not) wrap around the circuit in the presence of periodic (open) boundary conditions. At the end of the circuit are additional parameterized single-qubit $R_Y(\theta)$ gates on each qubit. (Bottom) Simulated noiseless VQE performance for preparing the ground state of the transverse field Ising chain with $h/J=1$ (A) and XXZ chain with $\Delta=-0.5$ (B) on $L=12$ qubits with periodic boundary conditions. The data shows lowest errors in ground state energy $1 - \left|\left(\langle H \rangle_{\theta} - \langle H \rangle_{g}\right)/\langle H \rangle_{g}\right|$ (red) and ground state fidelity $1 - |\langle \psi(\theta) | \psi_g \rangle |^2$ (blue) achieved by optimizing the checkerboard ansatz with $d=1,\dots , 12$ layers (increasing darkness), using the sequential least-squares programming (SLSQP) classical optimizer~\cite{kraft1988software} up to a given number of cost function evaluations. The curves are averaged over 10 simulated VQE instances with random parameter initialization and are shaded within $\pm 1$ standard error of the mean.}
   \label{fig:vqe_analysis}
\end{figure*}

We use a variational quantum circuit to prepare the ground states of the transverse field Ising, and XXZ chains.
In particular, we classically simulate a variational quantum eigensolver (VQE) to optimize circuit parameters, which we then use for state preparation in the experiment. 

VQE \cite{Peruzzo_2014} is a hybrid quantum-classical algorithm that classically optimizes parameters $\mathbf{\theta}$ of a variational ansatz $\vert\psi(\mathbf{\theta})\rangle = U(\mathbf{\theta})\vert \psi_0 \rangle$, with the objective of minimizing the cost function given by the expectation value of the Hamiltonian $\langle H \rangle_\mathbf{\theta}$. The unitary $U(\mathbf{\theta})$ is implemented by a parameterized quantum circuit. We initialize the algorithm from the $L$-qubit equal superposition state $\vert \psi_0 \rangle = \vert + \rangle^{\otimes L}$ and implement $d$ layers of the checkerboard ansatz $U^{(d)}$, as demonstrated for 6 qubits in Fig.~\ref{fig:vqe_analysis}. Our choice of circuit ansatz aligns with that of Bravo-Prieto et al. \cite{Bravo_Prieto_2020}, who observed that the critical number of ansatz layers required to accurately capture the energy and entanglement entropy of the ground states of the transverse field Ising and XXZ chain scales linearly with system size. 

Indeed, without experimental noise, VQE with $d=L=12$ ansatz layers is able to achieve sub-0.1\% errors in ground state energy and ground state fidelity of the transverse field Ising chain, and sub-1\% errors in ground state energy ground state fidelity of the XXZ chain, as seen in Fig.~\ref{fig:vqe_analysis}. Note that while increasing the number of layers $d$ improves the expressivity of the checkerboard ansatz, it also increases the difficulty of the variational optimization, since the number of ansatz parameters scales as $O(dL)$. Fig.~\ref{fig:vqe_analysis} shows that achieving 99\% accuracy requires hundreds and thousands of cost function evaluations by the optimizer.

\subsection{Error mitigation}
\label{SI:error mitigation}

Probabilistic error cancellation (PEC) is an error mitigation technique introduced in Ref.~\cite{temme2017error} (see also Refs.~\cite{endo2018practical,endo2021hybrid}). It can be used to calculate an effective expectation value $\langle A_N\rangle$ over $N$ noisy circuits that converges to the noise-free expectation value $\langle A\rangle$ for the corresponding observable $A$.
In particular, given an initial state $\rho_0$ and a unitary operator $\mathcal{U}$, we are interested in computing $\langle A \rangle = \Tr[A \mathcal{U}(\rho_0)]$. When the system is affected by noise we may only have access to a noisy operator $\tilde{\mathcal{U}} = \Lambda\circ\mathcal{U}$ that can be modeled as the ideal unitary $\mathcal{U}$ followed by some noise channel $\Lambda$. Direct measurement of the observable after applying the noisy gate $\tilde{\mathcal{U}}$ would then give $\langle \tilde{A}\rangle = \Tr[A\tilde{\mathcal{U}}(\rho_0)]$. If we have access to the inverse noise map $\Lambda^{-1}$ and could apply it after the noisy operation, we would be able to evaluate $\Tr[A\Lambda^{-1}\tilde{\mathcal{U}}(\rho_0)]$, which reduces to $\Tr[A\mathcal{U}(\rho_0)]$ and thus yields the ideal value $\langle A\rangle$. Unfortunately, the inverse noise map is generally an unphysical operation and therefore cannot be applied directly. This is where probabilistic error cancellation comes
in. Suppose we can write $\Lambda^{-1} = \sum_i \alpha_i \mathcal{V}_i$, where $\{\mathcal{V}_i\}$ is a set of physical operations that can be applied. The approach starts by rewriting $\Lambda^{-1}$ as:
\[
\Lambda^{-1} = \gamma \sum_i \sigma_i p_i \mathcal{V}_i,
\]
where $\gamma = \sum_i \vert \alpha_i\vert$, $\sigma_i = \mbox{sign}(\alpha_i)$, and $p_i = \vert \alpha_i\vert / \gamma$ such that $p$ is a probability distribution. Linearity of the trace means that we therefore have
\[
\Tr[A\Lambda^{-1}\tilde{\mathcal{U}}(\rho_0)]
= \sum_i p_i \left(\gamma\sigma_i \Tr[A\mathcal{V}_i\tilde{U}(\rho_0)]\right).
\]
For a given maximum number of circuits instances $N$ the algorithm would proceed by sampling $i_{\ell}$ for $\ell\in [1,N]$ according to the probabilities in $p$ and measure $a_{\ell} = \Tr[A\mathcal{V}_{i_{\ell}}\tilde{U}(\rho_0)]$. The empirical expectation value is then given by
\[
\langle A_N\rangle = \frac{\gamma}{N}\sum_{\ell=1}^N \sigma_{i_{\ell}} a_{\ell},
\]
where, notably, the signs $\sigma_{i_{\ell}}$ and the scaling factor $\gamma$ are applied in classical post-processing. The global scaling factor $\gamma$ increases the variance of the estimator. Compensating for this results in a sampling overhead by a factor $\gamma^2$.

\subsubsection*{Learning and inverting the noise channel}

For PEC to work we need an accurate model of the noise channel $\Lambda$, a scalable way of decomposing the inverse noise map $\Lambda^{-1}$, and an efficient way of sampling its decomposition in terms of operators $\mathcal{V}$. There are many protocols we could use to learn the noise channel. For example, gate-set tomography~\cite{greenbaum2015introduction, nielsen2021gate} could be used to fully characterize $\mathcal{\tilde{U}}$. However, this approach is not scalable for large systems. A more efficient approach for learning entire layers of two-qubit operations was introduced in Ref.~\cite{berg2022probabilistic}, which assumes a structured Pauli noise channel generated that can be written as a series of simple (and commuting) Pauli channels of the form $\Lambda_{i}(\rho) = \beta_i\rho + (1-\beta_i)P_i\rho P_i^{\dag}$, where $P_i$ is some one- or two-local Pauli operator. Each channel is easily inverted as $\Lambda_{i}^{i}(\rho) = \gamma_i(\beta_i\rho - (1-\beta_i)P_i\rho P_i)$, with scaling factor $\gamma_i = 1/(2\beta_i-1)$. Since the channels commute we could simply apply quasi-probabilistic sample to each of them, sampling the identity operator ($\sigma_i = 1$) with probability $\beta_i$ and a Pauli operator $\mathcal{V}_i(\rho) = P_i\rho P_i^{\dag}$ ($\sigma_i = -1$) with probability $(1-\beta_i)$. The operators, signs, and scaling factors inserted to the circuit could all be combined (see Ref.~\cite{berg2022probabilistic} for more details).

\subsubsection*{Experiment details}
\begin{figure*}[t]
\setlength{\tabcolsep}{10pt}
\centering
\includegraphics[width=1.0\textwidth]{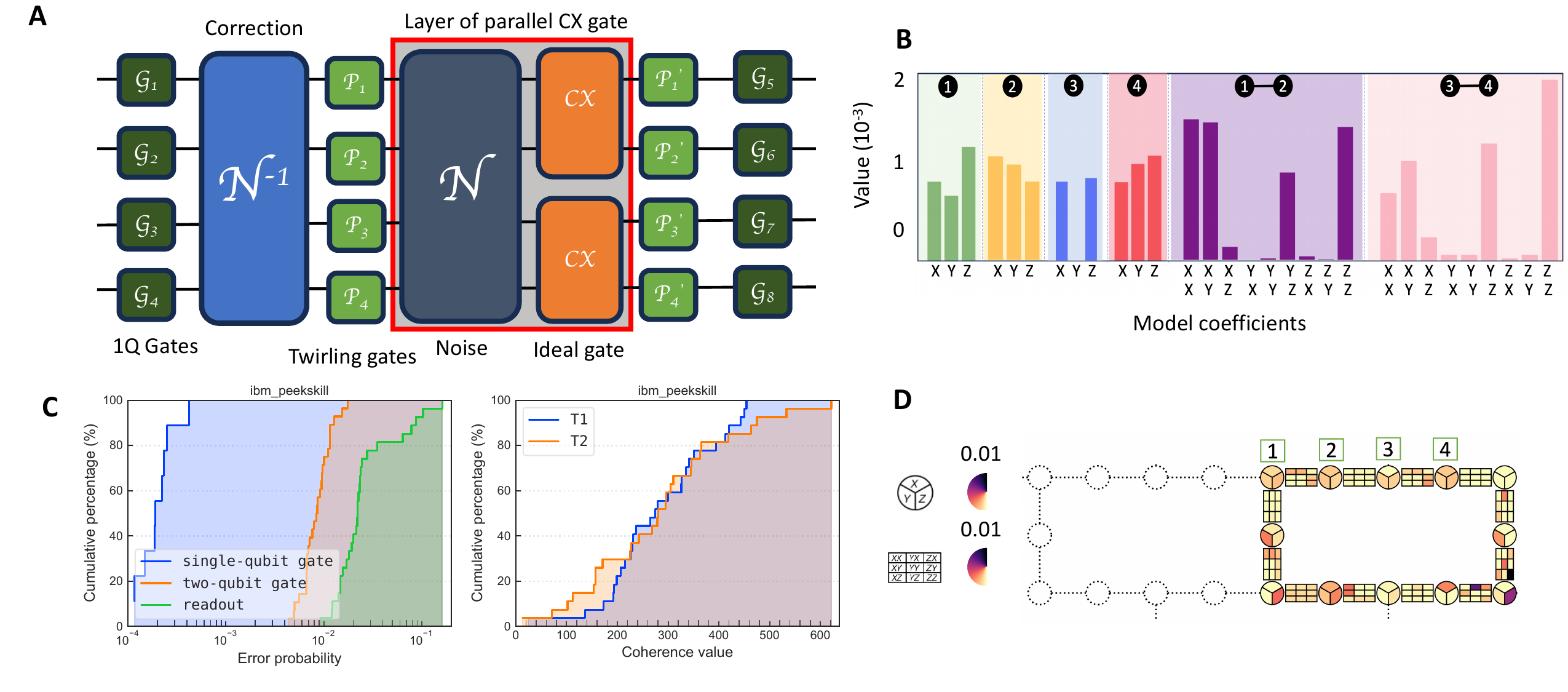}
\caption{\textbf{Experimental implementation of Probabilistic Error Cancellation}
(A) An example $4$ qubit template circuit for implementing PEC. Circuit consists of two qubit layers that we learn, twirling gates, and corrections.  (B) Displaying the model coefficients of the sparse Pauli Lindblad model over 4 qubits. (C) 
Empirical cumulative distribution functions (CDFs) of coherence values (T1 and T2) and gate error rates for one IBM quantum device: \textit{ibm\_peekskill}. The plots display the cumulative percentage of qubits with given T1 (blue) and T2 (orange) coherence values on the right and single-qubit gate errors (blue), two-qubit gate errors (orange) and readout error (green) on the left. The x-axis represents the coherence value (or error rates), while the y-axis shows the cumulative percentage of qubits. (D) Visualizing the sparse Pauli noise model generator coefficients on a $27$ qubit IBM Falcon processor. The circles denote the qubits with 3 internal wedges visualizing single X, Y, and Z Lindbladian coefficients and the two-body coefficients represent $3\times 3$ matrices related to different possible pairs. 
}
\label{fig-supp:experimental details}
\end{figure*}
For our problem, we learn the noise associated with two different layers of entangling operations present in the VQE, even layer (two qubit pairs [0, 1], [2, 3], [4, 5]) and odd layer (two qubit pairs [1, 2], [3, 4]) on a linear chain of qubits. In Fig.~\ref{fig-supp:experimental details} we summerize the experimental implementation details of PEC for one odd layer of CNOTs. We find $\overline{\gamma}$ ($\gamma$ per qubit) for these layers are 1.04 and 1.06. 

\subsection{Post-processing and error analysis for PEC}
\label{SI:Boostrapping}

\begin{figure}[t]
\setlength{\tabcolsep}{10pt}
\includegraphics[width=0.98\linewidth]{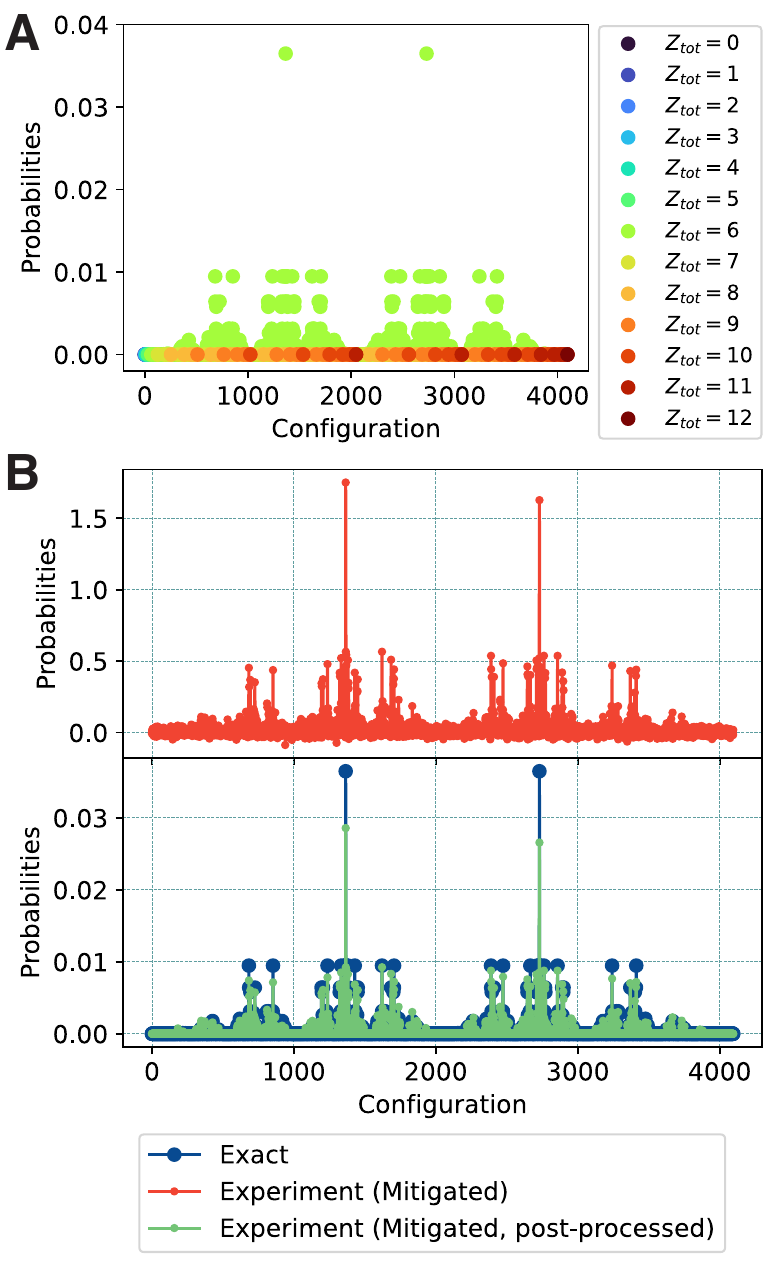}
\caption{\textbf{Post-processing of bitstring probabilities from PEC}. (A) Distribution of bitstring probabilities for the critical XXZ model ($\Delta=-0.5$) with periodic boundary conditions in $\sigma^z$ basis, computed via exact diagonalization (ED) for system size $L=12$. The bitstring states are eigenstates of the total magnetization $Z_\text{tot}$ and are colored according to their corresponding eigenvalue. Notably, $Z_\text{tot}$ is also conserved for the $U(1)$-symmetric XXZ model, and only the $Z_\text{tot}=L/2$ sector appears in the ground state. (B) Distribution of bitstring probabilities obtained from PEC (red) and the post-processing procedure (green), both averaged over bootstrap instances. As outlined in the text, for each bootstrapped dataset, the post-processing step nullifies probabilities of all bitstrings outside the $Z_\text{tot}=L/2$ sector, as well as probabilities below a cutoff in the $Z_\text{tot}=L/2$ sector. (The cutoff is determined by the root-mean-square magnitude of negative probabilities in the $Z_\text{tot}=L/2$ sector, which is informed by the PEC standard deviation.) The resulting non-negative set of probabilities are then renormalized. Comparison with the exact bitstring probabilities from panel A (blue) shows good agreement with the post-processed data.
}
\label{fig:postprocessing}
\end{figure}

The PEC method results in unnormalized probabilities, where some of the smallest probabilities take negative values within the standard deviation associated with the method. In this section, we provide details on how we employ bootstrapping to perform error analysis on the mitigated data and how we leverage symmetries of the quantum spin models to improve post-processing of the bitstring probabilities.

To report the accuracy of our observable estimates, we employ bootstrapping on the mitigated bitstring probabilities, which involves resampling the original data set to generate multiple simulated datasets. Through this process, we can access uncertainties by examining the standard deviation across the bootstrapped datasets. In our study, we execute 10,000 different mitigated circuits, each comprising 200 shots. We then resample 20 bootstrapped datasets of 1,000 mitigated instances each. Finally, we independently compute the target observables and quantities for each of the 20 bootstrapped datasets (see next paragraph), and report the mean and standard deviation across these datasets. 

We resort to post-processing in order to obtain a non-negative, normalized bitstring probability distribution for each bootstrapped dataset, see example procedure in Fig.~\ref{fig:postprocessing}. In particular, we utilize the symmetry of the underlying Hamiltonian in order to eliminate certain symmetry sectors that should not appear in the ground state. Namely, we remove the odd-parity (odd $Z_\text{tot}$) sectors for the TFI model in $\sigma^x$-basis and for the XXZ model in $\sigma^x$-basis, and only consider the $Z_\text{tot}=L/2$ sector for the XXZ model in $\sigma^z$-basis. Consequently, the magnitude of negative probabilities in the remaining symmetry sectors are significantly smaller as they are less biased to zero. We then compute the largest-amplitude negative probability in each bootstrapped dataset and set the probability cutoff $p_{\text{cutoff}}$ as the root-mean-square of these values across bootstrapped datasets. Setting all probabilities $p<p_{\text{cutoff}}$ to be zero, we then renormalize the probability distribution within each bootstrapped dataset and compute Shannon-R\'enyi entropies for subsystem configuration probabilities, as outlined in the main text. Then, we fit the functional form for $I_n(l,L-l)$ defined in Eq.~\ref{eqn:mutual_l} of the main text for even subsystem lengths $l\mod 2 = 0$, and extract the central charge for each bootstrapped dataset.

Finally, we report the central charge as the mean value across bootstrapped datasets. We report the error bar as the sum in quadratures of the random and systematic errors: $\sigma=\sqrt{\sigma_{\text{rand}}^2 + \sigma_{\text{sys}}^2}$. The random error $\sigma_{\text{rand}}$, which is calculated as the standard deviation of central charge values across bootstrap estimates, includes shot noise as well as the increased variance associated with PEC. The systematic error $\sigma_{\text{sys}}$ captures the error on the parameter estimates obtained from fitting the desired functional form Eq.~\ref{eqn:mutual_l} (we use the root-mean-square of the fit errors across bootstrapped datasets). In Table~~\ref{tab:SI_I}, we reported the value of central charge and related errors for TFI at critical point ($h/J$=1). The values obtained from fitting entanglement entropy in Eq.~\ref{eqn:CFT-EE-PBC-main} of the main text for various values of Renyi entropy $n$. The best value is related with $n=3$ indicating $c=0.49\pm 0.13$ compared to predicted theoretical value of $c=0.5$. 

\begin{table}[ht]
\centering 
\begin{tabular}{cc|cc}
\hline\hline
$n$ & $c$ &$n$ & $c$   \\ [0.5ex]
\hline
\\[-7pt]
1\,\,\,\,\, &0.59$\pm$0.10 &6\,\,\,\,\, &0.46$\pm$0.12\\[0.75ex] 
2\,\,\,\,\, &0.54$\pm$0.14 &7\,\,\,\,\,  &0.46$\pm$0.12 \\[0.75ex] 
3\,\,\,\,\, &0.50$\pm$0.13 &8\,\,\,\,\, &0.46$\pm$0.12\\[0.75ex] 
4\,\,\,\,\, &0.47$\pm$0.13 &9\,\,\,\,\, &0.45$\pm$0.12 \\[0.75ex] 
5\,\,\,\,\, &0.48$\pm$0.12. &10\,\,\,\,\, &0.45$\pm$0.12\\[0.75ex]  

\hline 
\end{tabular}
\caption{Values of central charge obtained from fitting different moments of entanglement entropy}
\label{tab:SI_I}
\end{table}

\subsection{Central charge for tricritical Ising model}
\label{SI:Tricritical}

Finally, we also examined a quantum spin chain recently introduced by~\cite{O'Brian2018} in the context of lattice supersymmetry, where the authors investigate the order-disorder coexistence in the tricritical Ising model. The Hamiltonian of the corresponding lattice model is defined as 
\begin{eqnarray}
\textbf{H}_{3}&=&-J\sum_{l=1}^{L}\sigma_l^z\sigma_{l+1}^z-h\sum_{l=1}^L\sigma_l^x\nonumber\\
&+& \lambda\sum_{l=1}^L(\sigma_l^x\sigma_{l+1}^z\sigma_{l+2}^z+\sigma_l^z\sigma_{l+1}^z\sigma_{l+2}^x)\;.
\label{tricritical}
\end{eqnarray}
While one recovers TFI for $\lambda=0$, the model remains in the Ising CFT universality class for $0\leq \lambda<\lambda_{TCI}$, right at $\lambda_{TCI}\approx 0.428$ there is a tricritical Ising (TCI) point. Moreover, the Hamiltonian is symmetric under $\sigma^z\rightarrow -\sigma^z$ and is also self-dual under Kramers-Wannier duality~\cite{O'Brian2018}. In a recent work, the conformal data has been extracted  from an accurate computation of  variational low-energy utilizing approach based on Bloch-state ansatz on periodic uniform matrix product state~\cite{Zou2018}. The numerical calculation predicts a value of $c=0.7$ for a system of $L=256$ at its critical point. In our experiment, possibly due to long-range interaction, the VQE algorithm has difficulty finding the ground state of the periodic chain of length $L=12$. However, we were able to prepare the ground state for the open boundary condition at $\lambda_{TCI}\approx 0.428$ of $L=10$ with a fidelity of 0.99, similar to what we shown in Supplementary material~\ref{SI:VQE}. Nonetheless, we observe stronger finite size effects here, both for exact diagonalization as well as experimental data leading to values of central charge of $c=1.0827$ and $c=1.03489$, respectively. This suggests that one should go to bigger system sizes for system with longer range interaction to obtain better estimate of central charge.

\end{document}